\documentclass[iop]{emulateapj}
\usepackage{natbib}
\usepackage{graphicx,color,rotating}
\usepackage{footnote,lineno}
\usepackage{ulem} 
\usepackage{graphicx,amssymb,amsmath,amsfonts,times,hyperref}
\usepackage{subfigure}
\usepackage{multirow}

\def \aap  {A\&A}

\def \apj  {ApJ}
\def \apjs  {ApJS}
\def \apjl  {ApJL}

\definecolor{darkgreen}{RGB}{51, 204, 51}

\begin{document}

\title{Deriving the contribution of blazars to the {\it Fermi}-LAT Extragalactic $\gamma$-ray background at $E>10$ GeV with efficiency corrections and photon statistics}  

\author{M. Di Mauro$^1$, S. Manconi$^{2,3}$,  H.-S. Zechlin$^{3}$,  M. Ajello$^{4}$, E. Charles$^{1}$, F. Donato$^{2,3}$}

\affil{$^1$W. W. Hansen Experimental Physics Laboratory, Kavli Institute for Particle Astrophysics and Cosmology, Department of Physics and SLAC National Accelerator Laboratory, Stanford University, Stanford, CA 94305, USA}
\affil{$^2$Department of Physics, University of Torino, via P. Giuria, 1, 10125 Torino, Italy}
\affil{$^3$Istituto Nazionale di Fisica Nucleare, via P. Giuria, 1, 10125 Torino, Italy}
\affil{$^4$Department of Physics and Astronomy, Clemson University, Kinard Lab of Physics, Clemson, SC 29634-0978, USA}


\begin{abstract}
The {\it Fermi} Large Area Telescope (LAT) Collaboration has recently released the Third Catalog of Hard {\it Fermi}-LAT Sources (3FHL), which contains 1556 
sources detected above 10 GeV with seven years of Pass 8 data.  
Building upon the 3FHL results, we investigate the flux distribution of sources at high Galactic latitudes ($|b| > 20^{\circ}$), which are mostly blazars. We use two complementary techniques: 1) a source-detection efficiency correction method and 2) an analysis of pixel photon count statistics with the 1-point probability distribution function (1pPDF).  With the first method, using realistic Monte Carlo simulations of the $\gamma$-ray sky, we calculate the efficiency of the LAT to detect point sources.  This enables us to find the intrinsic source count distribution at photon fluxes down to $7.5\times10^{-12}$ ph cm$^{-2}$s$^{-1}$.   With this method we detect a flux break at $(3.5\pm0.4) \times 10^{-11} \mathrm{ph}\,\mathrm{cm}^{-2}\mathrm{s}^{-1}$ with a significance of at least $5.4 \sigma$. The power-law indexes of the source count distribution above and below the break are $2.09\pm0.04$ and $1.07\pm0.27$, respectively. 
This result is confirmed with the 1pPDF method, which has a sensitivity reach of $\sim10^{-11}$ ph cm$^{-2}$s$^{-1}$. Integrating the derived source count distribution above the sensitivity of our analysis, we find that $(42\pm8)\%$ of the extragalactic $\gamma$-ray background originates from blazars.
\end{abstract}

\maketitle

\section{Introduction}
The Large Area Telescope (LAT) onboard the {\it Fermi Gamma-ray Space Telescope} has revolutionized our understanding of the $\gamma$-ray sky.
An important achievement of the LAT was the measurement of the extragalactic $\gamma$-ray background (EGB) with unprecedented precision~\citep[from 100 MeV to 820 GeV at Galactic latitudes $|b|>20^{\circ}$, see][]{Ackermann:2014usa}. 
The EGB is composed of the emission from individual sources detected by the LAT as well as the isotropic diffuse $\gamma$-ray background (IGRB). The IGRB describes a component which is isotropic on angular scales larger than $\sim 1$ degree and whose composition is dominated by unresolved sources, i.e., sources that are not individually detected by the LAT.
The IGRB has been successfully interpreted as being composed of $\gamma$-ray emission from blazars, radio galaxies and star-forming galaxies \citep[see, e.g.,][]{Ajello:2011zi,DiMauro:2013zfa,DiMauro:2013xta,DiMauro:2014wha,DiMauro:2015tfa,Ajello:2015mfa}. These analyses are based on studies of the $\gamma$-ray population of blazars in {\it Fermi}-LAT catalogs and on correlations between the $\gamma$-ray and radio (for radio galaxies) or infrared emission (for star-forming galaxies), as the $\gamma$-ray sample of detected sources is limited for radio galaxies and star-forming galaxies.
The estimate of the contributions of the various source classes to the EGB relies on extrapolations of cataloged sources to fluxes below the sensitivity of the LAT, and on correlations between $\gamma$ rays and other wavelengths.  Such extrapolations may suffer significant uncertainties \citep[see, e.g.,][]{2012ApJ...755..164A,DiMauro:2013xta}.
The total contribution of blazars and radio galaxies, which are both radio loud Active Galactic Nuclei (AGN), and star-forming galaxies is known with an uncertainty between a factor of 2 and 3 \citep[see, e.g.,][]{DiMauro:2015tfa,Ajello:2015mfa}. 
Other mechanisms, such as $\gamma$ rays originating in AGN-driven shocks propagating through galaxies \citep[see, e.g.,][]{Lamastra:2017iyo}, have been invoked as alternative interpretations \citep[see, e.g.,][for a recent review]{Fornasa:2015qua}.

A technique that is less dependent on extrapolation is the calculation of the LAT efficiency for the detection of sources that can be used to correct the source count distribution, i.e.~the flux distribution, of cataloged sources.  The LAT Collaboration has used this method to measure the contribution of blazars to the extragalactic $\gamma$-ray sky at $E>50$ GeV \citep{TheFermi-LAT:2015ykq}, finding $86^{+16}_{-14}\%$. This result translates into tight constraints on the contribution of star-forming galaxies to the high-energy part of the EGB and to the IceCube astrophysical neutrino spectrum \citep[see, e.g.,][]{Bechtol:2015uqb}. 

Another method, which has been successfully applied to constrain the source count distribution, is the analysis of pixel photon count statistics. In particular, the 1-point probability distribution function (1pPDF) of photon count maps has been demonstrated to be more sensitive than the efficiency correction method. This result is unsurprising given that it includes information from unresolved sources contributing only very few (even single) photons per pixel \citep[see, e.g.,][]{Malyshev:2011zi,TheFermi-LAT:2015ykq,Lisanti:2016jub,Zechlin:2015wdz,Zechlin:2016pme}.

Recently, the {\it Fermi}-LAT Collaboration has released The Third Catalog of Hard {\it Fermi}-LAT Sources (3FHL), a new catalog of 1556 sources detected between $10$ and $2000$\,GeV with 7 years of data processed with the Pass 8 event-level analysis \citep{TheFermi-LAT:2017pvy}.  
The significant improvement in acceptance and angular resolution of Pass 8 \citep{Atwood:2013dra} and the large set of $\gamma$-ray data available after 7 years enabled an increase by a factor of three of the number of sources detected with the 3FHL with respect to the 1FHL. The 1FHL is the previous {\it Fermi} catalog of sources detected in the same energy range but with three years of operations \citep{Ackermann:2013fwa}. Among the 3FHL sources detected at $|b|>20^{\circ}$ (1120), $90\%$ are extragalactic ($89\%$ are blazars), $9\%$ are unassociated and $1\%$ are pulsars. 
If the unassociated sources follow the same ratio between the number of blazar and non-blazar sources as for the associated sources ($89/91$ for blazars and $2/91$ for non-blazars), we expect that $\sim 98\%$ of all high-Galactic-latitude 3FHL sources are blazars.
In short, the $\gamma$-ray sky above 10 GeV and at $|b|>20^{\circ}$ is dominated by extragalactic sources, and, in particular, by blazars.

In this paper, we employ both the efficiency correction method and the 1pPDF method to find the source count distribution above 10 GeV and at $|b|>20^{\circ}$.   We use the results to derive the contribution from point sources to the EGB at $E>10$ GeV.

The paper is organized as follows: Sec.~\ref{sec:datasel} contains the data selections and background models that we employ in our analysis; in Sec.~\ref{sec:effmethod} we present the results for the efficiency correction method, in Sec.~\ref{sec:1ppdf} we report the results derived with the 1pPDF and finally in Sec.~\ref{sec:lf} we compare the source count distribution derived with our analysis with the prediction for blazar luminosity function.

\section{Data selection and background models}
\label{sec:datasel}
We analyze seven years of Pass 8 data, from 2008 August 4 to 2015 August 4, i.e. this is the same data set as used for the 3FHL.
We select $\gamma$-ray events at Galactic latitudes $|b|>20^{\circ}$ in the energy range $E=[10,1000]$ GeV, passing standard data quality selection criteria.
For the efficiency correction method we consider events belonging to the Pass~8 {\tt SOURCE} event class, and use the corresponding 
instrument response functions (IRFs) {\tt P8R2\_SOURCE\_V6}, since we are interested in point source detection. 
The data selection for the 1pPDF analysis requires more stringent event cuts to reduce systematic uncertainties, in particular those related to point-spread function (PSF) smoothing and the effects of residual cosmic-ray contamination of the $\gamma$-ray event sample. Thus, for the 1pPDF analysis, we compare individual analyses of data for the {\tt SOURCE}, {\tt CLEAN}, and {\tt ULTRACLEANVETO} (UCV) event selections. Correspondingly, we use the {\tt P8R2\_SOURCE\_V6}, {\tt P8R2\_CLEAN\_V6}, and {\tt P8R2\_ULTRACLEANVETO\_V6} IRFs. PSF smoothing is minimized by restricting data selection to events belonging to the {\tt PSF3} quartile. For both analysis methods we select events with a maximum zenith angle of $105^{\circ}$, in order to minimize contamination from the Earth's atmosphere.\footnote{See \url{http://fermi.gsfc.nasa.gov/ssc/data/analysis}.}  We note that no rocking angle cut was applied.  For the 1pPDF analysis, the counts data are pixelized using the HEALPix pixelization~\citep{2005ApJ...622..759G} with order 7 and 8 (corresponding to $N_\mathrm{side}=128\,(256)$ and angular resolution of 0.46\,(0.23)\,deg). Tab.~\ref{tab:analysis} reports some key points of the two analyses.
The different selection criteria applied to efficiency correction and 1pPDF give a set of photons with a density of about $8.7$ ph/deg$^2$ and $1.1$ ph/deg$^2$ (for the UCV selection), respectively. Correspondingly, for the UCV selection the flux threshold for a source to be detected with on average one photon increases by a factor of about 7 with respect to the selection cuts used for the efficiency correction method.

We employ two different interstellar emission models (IEMs), in order to estimate the systematic uncertainties introduced by the choice of IEM. 
The first IEM is the {\tt gll\_iem\_v06.fits} template released with Pass 8 data \citep{Acero:2016qlg}.
This is the model routinely used in Pass~8 analyses and we refer to it as the {\it official} model (Off.).
The second template represents the ``Sample'' model of the Pass~8 analysis of the Galactic Center \citep{FermiPass8GC}. 
This model, which we call the {\it alternate} model (Alt.), contains newer IEMs, for example with a data-driven template for the Fermi Bubbles \citep{2010ApJ...724.1044S,Fermi-LAT:2014sfa} as well as an additional population of electrons used in modeling the central molecular zone.   Many of the other components of the Alt. model were computed with the GALPROP Galactic CR propagation code\footnote{See \url{http://galprop.stanford.edu}.}.

We also include the standard template for the isotropic emission ({\tt iso\_P8R2\_SOURCE\_V6\_v06.txt})~\footnote{For descriptions of these templates, see
  \url{http://fermi.gsfc.nasa.gov/ssc/data/access/lat/BackgroundModels.html}.}
that includes the IGRB and the so-called CR background component made of CRs misclassified in the LAT and $\gamma$ rays from the Earth’s atmosphere that have directional reconstruction errors sufficient to bypass the zenith angle veto \citep[see, e.g.,][]{Ackermann:2014usa}.

\begin{table*}
\center
\begin{tabular}{cccccc}
Analysis & IRF        & PSF type & $N_{\rm{events}}$        & PSF size [deg]  &   $F_{1\rm{ph}}$ [ph cm$^{-2}$\,s$^{-1}$]    \\
\hline
$\omega$	&  {\tt P8R2\_SOURCE\_V6}	 &  0,1,2,3	 & 235,399  & 0.16  &  $2.7\times 10^{-12}$\\
\hline
\multirow{3}{*}{1pPDF}	& {\tt P8R2\_SOURCE\_V6}	 & 3 & 59,319 & 0.08 & $1.3\times 10^{-11}$\\
	&  {\tt P8R2\_CLEAN\_V6} &  3	 &  46,527 &  0.08  &  $1.4\times 10^{-11}$\\
	&  {\tt P8R2\_ULTRACLEANVETO\_V6} &  3	 &   31,098 &  0.08  &  $1.8\times 10^{-11}$\\
\end{tabular}
\caption{The table reports details of the two analysis setups (efficiency correction first row and 1pPDF second row). The columns list the IRFs, the PSF types, the number of events in the ROI, the size of the PSF at 10\,GeV derived with {\tt gtpsf} tool, and the flux of a source detected with one photon.}
\label{tab:analysis}
\end{table*}

\section{Efficiency Correction Method}
\label{sec:effmethod}
The 3FHL source count distribution at $|b|>20^{\circ}$, $dN/dS$, follows a power law (PL) for $S >5\times10^{-11}$ ph cm$^{-2}$ s$^{-1}$, 
where $S$ is the integral photon flux for $E=[10,1000]$ GeV in units of ph cm$^{-2}$s$^{-1}$ \citep{TheFermi-LAT:2017pvy}. 
Below this flux, the observed $dN/dS$ drops quickly, owing to the difficulty in detecting fainter sources with the LAT.  In this section, we derive the LAT efficiency to detect a source with a given photon flux $S$, and we will use it to correct the $dN/dS$ distribution of the 3FHL catalog.

\subsection{Analysis pipeline}
\label{sec:pipeff}
We have implemented an analysis pipeline using {\tt FermiPy}, a Python package that automates analyses with the {\it Fermi} Science Tools \citep{2017arXiv170709551W}\footnote{See \url{http://fermipy.readthedocs.io/en/latest/}.}.
Specifically, {\tt FermiPy} tools are employed to 1) generate simulations of the $\gamma$-ray sky, 2) detect point sources, and 3) calculate the characteristics of their spectral energy distributions (SED).   
In general, the same analysis is applied to real and simulated data.  

We subdivide the sky at $|b|>20^{\circ}$ into 144 smaller regions of interest (ROIs)
of $22^{\circ}\times22^{\circ}$, each with an overlap of $4^{\circ}$ between
adjacent ROIs. This overlap is included so that sources near the edge of a ROI are well
contained in the adjacent ROIs. When a source is detected in more than one ROI we keep the 
one that is the closest to the center of its ROI.
We analyze each ROI separately, because considering the entire sky would imply several thousand free
parameters, making analyses with the {\it Fermi} Science Tools infeasible.  
In each ROI, we bin the data with a pixel size of $0.06^{\circ}$ and 12 energy bins per decade.

For each ROI our initial model includes only the IEM and the isotropic template.
The IEM has in our analysis the normalization and slope free to vary while for the isotropic template only the normalization is a free parameter.
Point sources are detected with an iterative process, where first sources with Test Statistics (TS)
 $>64$ are extracted and added to the model.
We use the likelihood ratio test to estimate the significance of source candidates.
The $TS$ is defined as twice the difference in log-likelihood between the null hypothesis (i.e., no source present) and the test hypothesis: $TS = 2 ( \log\mathcal{L}_{\rm test} - \log\mathcal{L}_{\rm null} )$ \citep{wilks1938}.
Subsequently, this procedure is repeated for sources with $TS>36$, 16, and 9{\footnote{$TS$ values of $TS=[9,16,20,25,36,64]$ approximately correspond to statistical significances in standard deviations of $[1.9,3.0,3.5,4.1,5.1,7.2]$.}.
After each step, a fit to the ROI is performed in order to derive the SED parameters of the sources. 
A PL shape is considered for all source SEDs.
At the end of this process all sources with $TS>9$ are found and included in the model.
The position of the sources is derived by {\tt FermiPy} making a $TS$ map and finding the peak of the $TS$ at the location of each source. This method gives also the $68\%$ error of the source position.
For more details on {\tt FermiPy} we refer to the Appendices of \cite{Fermi-LAT:2017yoi}. 

The source algorithms employed in the 3FHL catalog are {\tt mr\_filter} and {\tt PGWave} that are based on wavelet analysis in the Poisson regime \citep{1997ApJ...483..350D,1998A&AS..128..397S}.

Since we use this pipeline to generate and analyze the simulations, we apply the same method to derive the list of sources detected in the real sky.  We conduct a series of checks, as explained in the following section, comparing our list of sources with the official 3FHL catalog in order to validate our analysis. 

\subsection{Sources detected in the real sky}
\label{sec:3FHLcat}
We create our source lists using the procedure described in Sec.~\ref{sec:pipeff}.
We make two versions of this list, using the Off. and the Alt. IEM respectivley, to investigate the effect of uncertainties of the Galactic diffuse emission modelling on the number and properties of the detected sources.
In Table~\ref{tab:3FHL} we summarize the results for $TS>16$ and $>25$.
We find a number of sources that is consistent within $2\sigma$ with the official 3FHL catalog.
We checked that the 3FHL sources that we do not detect with our analysis are faint sources with $TS$ near the threshold.  
The difference in the detected sources between our analysis and the 3FHL is thus attributed to threshold effects.
We also compare the mean and RMS of the photon index ($\Gamma$) distribution of detected sources with $TS>25$, and find that is $2.57\pm0.65$ for our catalog with the Off. IEM and $2.62\pm0.75$ for the 3FHL catalog.
The choice of the Alt. IEM does not affect the number and SED properties of detected sources (see Table~\ref{tab:3FHL}).

\begin{table}[t]
\center
\begin{tabular}{c|cc|c|cc}
$TS>25$ & $N_{\rm{det}}$        & $\Gamma$ & $TS>16$  & $N_{\rm{det}}$        & $\Gamma$     \\
\hline
3FHL	&	986 &	$2.60\pm0.72$ &  3FHL    &	 &	  \\
Off. IEM	&	929 &	$2.57\pm0.65$ &  Off. IEM    &	1490&	$2.52\pm0.71$  \\
Alt. IEM	&	930 &	$2.57\pm0.65$ &  Alt. IEM    &	1496 &	$2.52\pm0.72$  \\
Simulations	&	935 &	$2.62\pm0.68$ &  Simulations    &	 1652 &	$2.53\pm0.77$  \\
\end{tabular}
\caption{Number of sources ($N_{\rm{det}}$) and photon index ($\Gamma$) distribution for the sources detected with $TS>25$ (left block) and $TS>16$ (right block) for the 3FHL catalog, for our analysis of the real sky with the Off. and Alt. IEMs, and for the simulations. The quoted errors are RMS values of the $\Gamma$ distribution. The space for the official 3FHL catalog for $TS>16$ is empty because the catalog has been derived for sources detected with $TS>25$ only.}
\label{tab:3FHL}
\end{table}

In Fig.~\ref{fig:3FHL} we show the comparison of photon fluxes for 3FHL catalog sources ($S_{\rm{3FHL}}$) with the list of sources detected in our analysis ($S$).
Each source detected in our analysis is associated with a source in the 3FHL catalog using the $95\%$ positional uncertainty as given in our analysis ($r_{95}$) and in the 3FHL catalog ($r_{95,\rm{3FHL}}$), i.e. if the angular distance between the sources is 
smaller than $\sqrt{{r_{95}}^2 + {r_{95,\rm{3FHL}}}^2}$.

Sources with a flux $S>10^{-10}$ ph cm$^{-2}$ s$^{-1}$ have a difference in photon flux within $10\%$. For fainter sources the differences reach at most the $20\%$ level.  However, the statistical errors on the measured $S$ are of the same order of these differences.  
In the official 3FHL catalog 986 sources have been detected at $|b|>20^{\circ}$ while in our pipeline we find 929 and 930 with the Off. and Alt. IEM. We detect with $TS>25$ the 91\% of 3FHL sources and we checked that this fraction increases to 99\% considering sources detected in our analysis with $TS>16$. 

In short, the number of detected sources in our analysis as well as the $S$ and $\Gamma$ distributions of those sources are compatible with the results presented in the 3FHL.  

Since we use our pipeline to find the efficiency, we apply the same pipeline also to derive the list of sources detected in the real sky, in order to be fully self-consistent.

\begin{figure}
	\centering
\includegraphics[width=1.03\columnwidth]{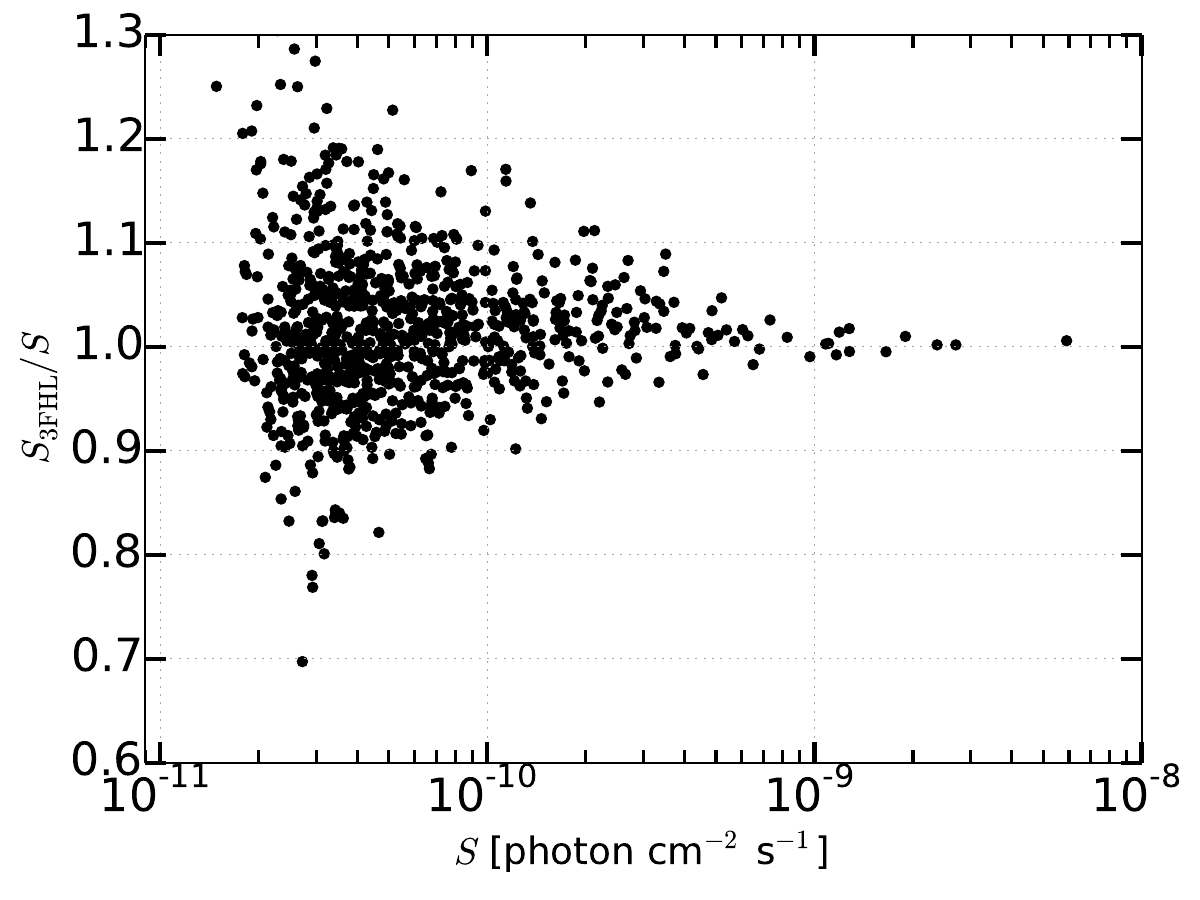}
\caption{Ratio between the photon flux of sources detected in the 3FHL catalog ($S_{\rm{3FHL}}$) and in our analysis ($S$).}
\label{fig:3FHL} 
\end{figure}

\subsection{Calibrating the IEM and isotropic emission}
\label{sec:enspectrum}
In order to derive the efficiency we need to create and analyze simulations of the $\gamma$-ray sky.
These simulations include the IEM, the isotropic template, and point sources with a flux drawn from a broken PL (BPL) shaped $dN/dS$ (see Eq.~\ref{eq:BPL}).  
If we use the IEM and isotropic templates directly as given, i.e. with an overall normalization equal to 1, we are not able to correctly reproduce the energy spectrum of the real sky.
Therefore, we first calibrate the normalization of the IEM and isotropic template.
We create simulations that we label as {\it energy-spectrum simulations}, including the IEM, the isotropic template, and flux from sources detected in the real sky with $TS>25$ in our analysis.
We perform {\it energy-spectrum simulations} dividing the sky at $|b|>20^{\circ}$ in 16 different ROIs, 
and we conduct a likelihood fit using Poisson statistics comparing the energy spectrum (number of photons per energy bin) of the {\it energy-spectrum simulations} in each ROI with the one of the real sky. 
In the likelihood fit, the normalizations of the IEM and isotropic template are free to vary.
We use fewer ROIs in this case because we only need to find the energy spectrum. 
As a result, we obtain that the best-fit normalization of the Off. (Alt.) IEM is $1.32\pm 0.06$ ($1.09\pm 0.04$), together with a normalization of the isotropic template of $0.81\pm 0.04$ ($0.46\pm 0.04$). In Fig.~\ref{fig:norm} we show the best-fit results for the Off. IEM in comparison with the energy spectrum of the {\it energy-spectrum simulations} and the real sky. The difference between the {\it energy-spectrum simulations} and the real sky is at most $10\%$ below 100 GeV, while above it can reach the 20\% level (where, however, the statistics of $\gamma$ rays is quite limited).

\begin{figure}
	\centering
\includegraphics[width=1.03\columnwidth]{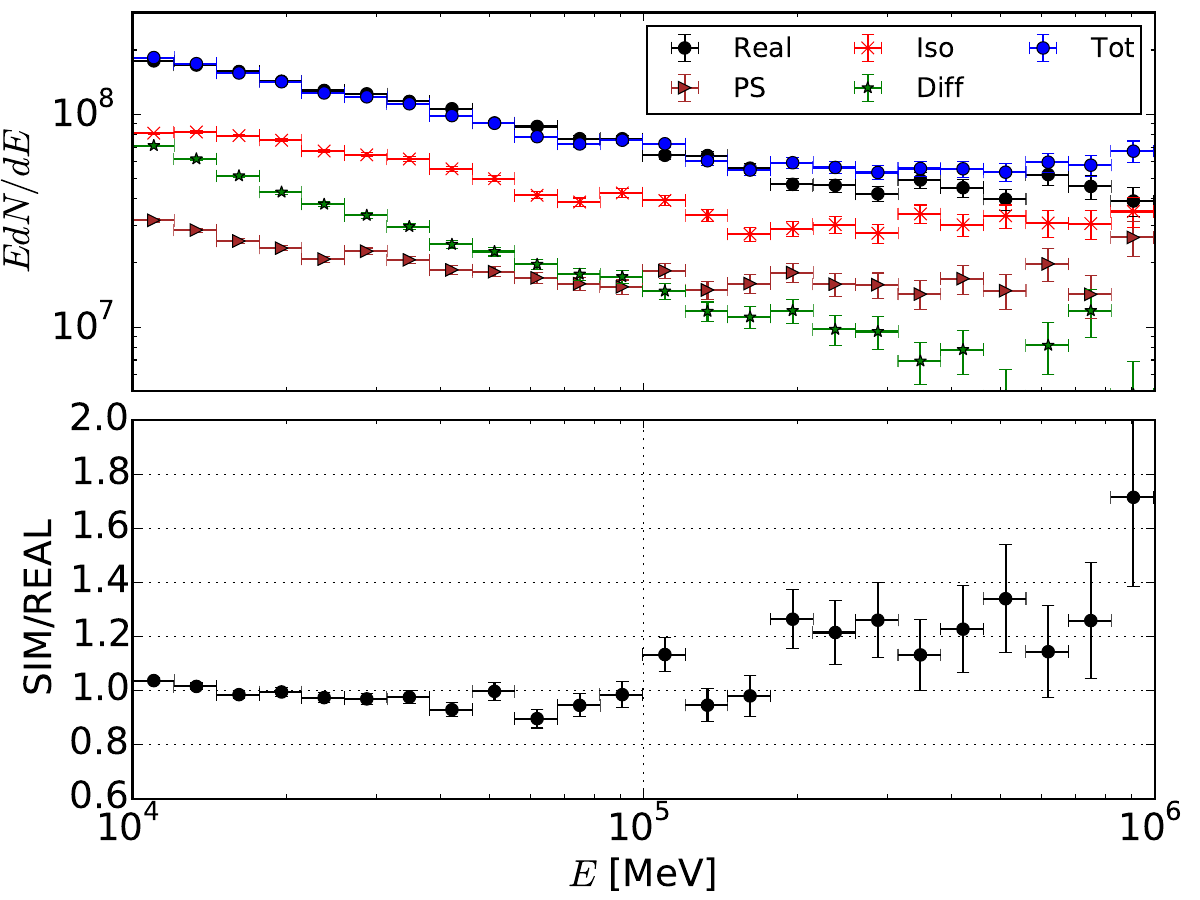}
\caption{Top panel: Energy spectrum of $\gamma$ rays (number of photons per energy bin) in the real sky (black points) and for the best-fit scenario of our {\it energy-spectrum simulations} (blue), together with the simulated Off. IEM (green), isotropic (red), and detected sources (brown). Bottom panel: ratio of the energy spectrum of {\it energy-spectrum simulations} and of the real sky. 
}
\label{fig:norm} 
\end{figure}

The normalizations of the IEM and isotropic component found before are valid if we consider cataloged sources, i.e. sources detected with $TS>25$.  
However, in our estimation of the efficiency we will consider the real source population including also the undetected ones. We thus need to change accordingly the isotropic template found before and reproduce correctly the real sky energy spectrum.
In order to account for this effect we use the following procedure.
We consider the total flux of sources detected in our list with $TS>25$ ($S_{\rm{PS}}$) and the flux of the isotropic template ($S_{\rm{Iso}}$). 
These two components together constitute the so-called EGB.
The level of EGB must be the same also when we produce our simulations and we include undetected sources.
Simulating sources from a flux below the threshold of the catalog has the effect of increasing the contribution of point sources and reducing the one from the isotropic emission.
Therefore, the total flux from simulated sources ($S^{\rm{sim}}_{\rm{PS}}$) and the isotropic template ($S^{\rm{sim}}_{\rm{Iso}}$) must satisfy the following rule: $S_{\rm{PS}} + S_{\rm{Iso}} = S^{\rm{sim}}_{\rm{PS}} + S^{\rm{sim}}_{\rm{Iso}}$. Using this conservation law, for a given shape of the source count distribution of simulated sources we can find the normalization to use in the simulations for the isotropic templates.
\begin{equation}
\rm{Norm}_{\rm{ISO}} = \frac{S_{\rm{PS}} + S_{\rm{ISO}} - S^{\rm{sim}}_{\rm{PS}}}{ S_{\rm{ISO}}}.
\label{eq:Isonorm}
\end{equation}
The value of $\rm{Norm}_{\rm{ISO}}$ depends on the shape of the $dN/dS$ used in our simulations, meaning that we calculate a value of $\rm{Norm}_{\rm{ISO}}$ for every $dN/dS$ used to draw the flux of sources.
Therefore, when we simulate sources with a given $dN/dS$ we normalize the isotropic emission with $\rm{Norm}_{\rm{ISO}}$. 

\subsection{Simulations to derive the efficiency}
\label{sec:sim}
In the simulations used to find the efficiency, that we label as {\it efficiency simulations}, we simulate sources using a BPL shape for the source count distribution:
\begin{eqnarray}
\label{eq:BPL}
\frac{dN}{dS} = K
\left\{
\begin{array}{rl}
& S^{-\gamma_1} S_{\rm{b}}^{\gamma_1-\gamma_2}, \quad
 S \leq S_{\rm{b}}, \\
& S^{-\gamma_2}, \quad
S >S_{\rm{b}},
\end{array}
\right.
\end{eqnarray}
where $\gamma_2$ and $\gamma_1$ are the slopes of the $dN/dS$ above and below the break flux $S_{\rm{b}}$.
We generate 5 simulations with $\gamma_2=2.20$, $\gamma_1=1.50$, and $S_{\rm{b}}=5\times 10^{-11}$ ph cm$^{-2}$ s$^{-1}$, since, as we will see, this shape of the source count distribution approximates the correct one. We draw fluxes for $S>7\times 10^{-13}$ ph cm$^{-2}$ s$^{-1}$, thus one order of magnitude below the faintest source detected in the real sky with $TS>16$.
In Sec.~\ref{sec:valid}, we also use a Log Parabola (LP) to parameterize $dN/dS$:
\begin{equation}
\label{eq:LP}
\frac{dN}{dS} = K \left(\frac{S}{S_0}\right)^{-\alpha+\beta \log{(\frac{S}{S_0})}},
\end{equation}
where $\alpha$ is the slope of the distribution and $\beta$ its curvature. For the IEM, we use the normalizations found in Sec.~\ref{sec:enspectrum}, ($1.32\pm 0.06$ for the Off. and $1.09\pm 0.04$ for the Alt.~IEM) 
while we renormalize the isotropic template using Eq.~\ref{eq:Isonorm}, obtaining 0.74 for the Off. and 0.40 for the Alt. IEM.
Finally, for each source we draw the photon index $\Gamma$ from a Gaussian distribution with mean 2.55 and standard deviation 0.40.
We will show that given this intrinsic distribution of indexes we recover the observed distribution of $\Gamma$ for detected sources in the real sky. 
The difference between the input and detected photon index distributions is given by detection biases \citep{2010ApJ...720..435A,TheFermi-LAT:2015ykq}.


We employ the same tools and pipeline as used for the real data to detect sources in simulated data.
In Tab.~\ref{tab:3FHL} we report the number of sources and the photon index distribution for sources detected with $TS>25$ and $TS>16$ from the {\it efficiency simulations}. $\Gamma$ and $N_{\rm{det}}$ (for $TS>25$) as found from the simulations are compatible with the sources detected in the real sky. 
On the other hand $N_{\rm{det}}$ for $TS>16$ in the simulations is slightly higher with respect to the real sky but this is not going to affect our results.
For each source, Fig.~\ref{fig:simdet} compares the flux drawn from the simulated $dN/dS$ ($S_{\rm{SIM}}$) with the flux measured from detecting the same source in the simulated data ($S$). A source detected in our analysis is associated with a simulated source if the position of the latter is within the $95\%$ positional error of the detected source.
The flux values are well compatible for sources with $S>10^{-10}$ ph cm$^{-2}$s$^{-1}$, while fainter sources show a bias that increases the flux of the detected source with respect to the flux with which it was actually simulated.
This is associated to the Eddington bias that describes the statistical fluctuations of sources from simulated flux to detected flux \citep{eddington1913}. Since the number of sources with a true flux (simulated flux) below the LAT detection threshold is larger than the number of sources with a flux above the threshold, many simulated sources are detected with a flux greater than the true flux.

\begin{figure}
	\centering
\includegraphics[width=1.03\columnwidth]{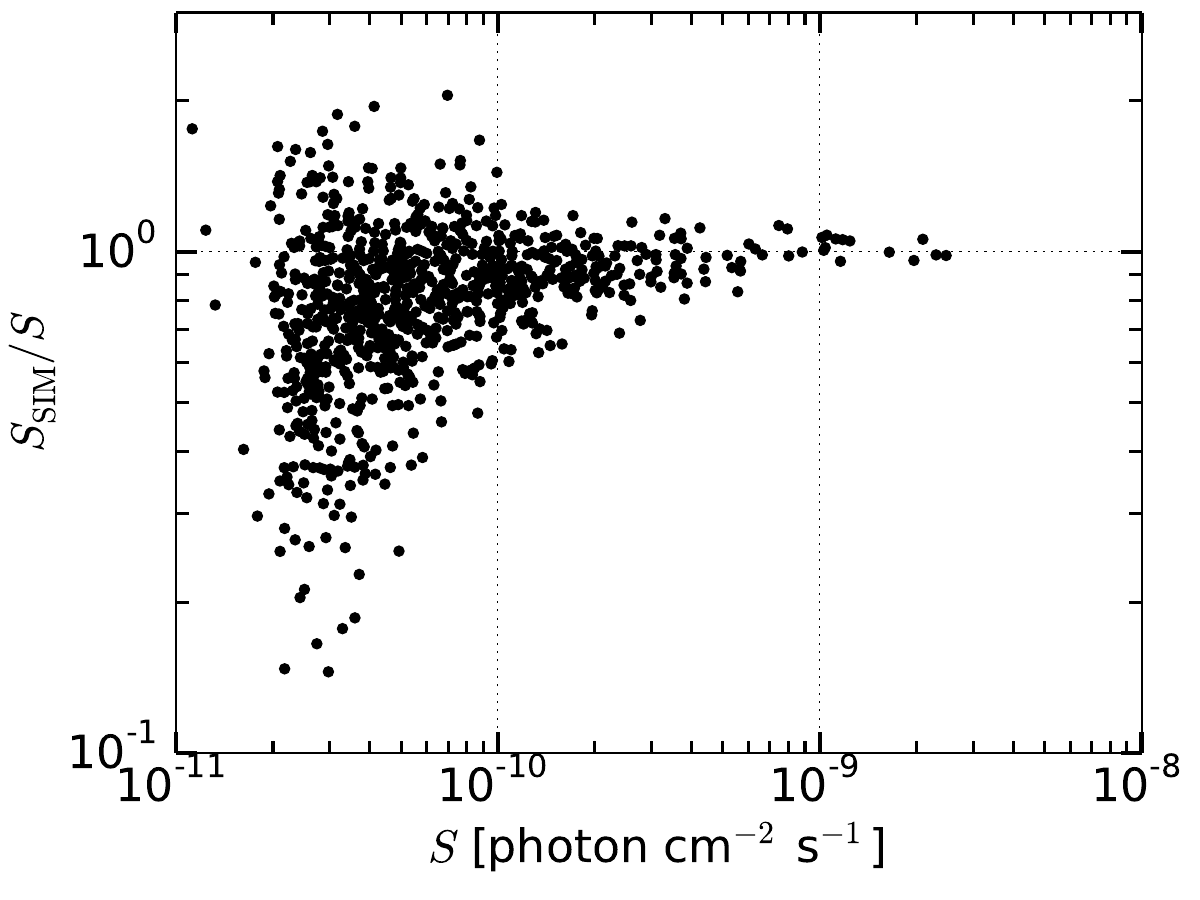}
\caption{Ratio between the true photon flux $S_\mathrm{SIM}$ used as simulation input and the measured photon flux $S$ as reconstructed from the simulated data. Each point corresponds to an individual source with measured photon flux $S$.}
\label{fig:simdet} 
\end{figure}

\subsection{Spurious sources}
\label{sec:spurious}
One of the main novelties of this paper is that we apply the efficiency correction method also to sources detected below the canonical value of $TS=25$ used by the {\it Fermi} Collaboration as a threshold to include a source in their catalogs.
Indeed, we consider also sources detected with $TS>20$ and $TS>16$ in order to constrain the source count distribution at fainter fluxes and with better statistics.
Lowering the $TS$ threshold makes it more likely to detect sources that are spurious, meaning that they are originated by statistical fluctuations of the IEM or isotropic emission and thus are not real.
A source detected with $TS=16$ at $E>10$ GeV originates on average from 4 photons \citep{TheFermi-LAT:2017pvy}. The isotropic and IEM components instead contribute 0 or 1 photons per pixel, meaning that we need a $4\sigma$ fluctuation and a p-value of $6.3 \times 10^{-5}$ to have a spurious source coming from these components and detected with $TS=16$. Since the sky is covered by a few million pixels, we expect to detect a few hundred spurious sources when considering $TS>16$. This rough estimation gives a sense of how many spurious sources may contribute to the list of sources detected with $TS>16$. From the same calculation we expect that the number of spurious sources detected $TS>25$ should be negligible.

We employ our simulations to derive the fraction of spurious sources with respect to the total number of detected sources as a function of the observed photon flux. This quantity is denoted by $\mathcal{R}$. This information will be used in the next section for the calculation of the efficiency.
We assume that a source is spurious if no simulated source exists within $95\%$ CL positional uncertainty.
Since we use the $95\%$ CL positional uncertainty we intrinsically miss $5\%$ of real sources that we label as spurious. We will account for this in our analysis.
We show this result in Fig.~\ref{fig:spurious}  for the sample of sources detected at $TS>16/20/25$ for the {\it efficiency simulations}. 
Considering $TS>16$ and $>20$, the fraction of spurious sources is constant at around $2\%$ level for sources detected with $S>4\times10^{-11}$ ph cm$^{-2}$ s$^{-1}$. This means that the number of spurious sources at these fluxes is compatible with zero. On the other hand, $\mathcal{R}$ increases for fainter sources, up to almost one for $S< 10^{-11}$ ph cm$^{-2}$ s$^{-1}$.
The effect of spurious sources for this source sample is thus negligible for sources detected with $S>4\times10^{-11}$ ph cm$^{-2}$ s$^{-1}$ but becomes important when we consider fainter sources. For sources detected with $TS>25$, $\mathcal{R}$ is at the same level or lower than $5\%$ for $S>2\times10^{-11}$ ph cm$^{-2}$ s$^{-1}$ and it is slightly larger, even if with large statistical uncertainties, below this flux. For this subsample of sources the contribution of spurious sources is thus negligible.

We checked that for the $99\%$ CL positional uncertainty we obtain on average $4\%$ fewer spurious sources, consistent with the chosen CL that is larger by $4\%$ with respect to the $95\%$ CL.

\begin{figure}
	\centering
\includegraphics[width=1.03\columnwidth]{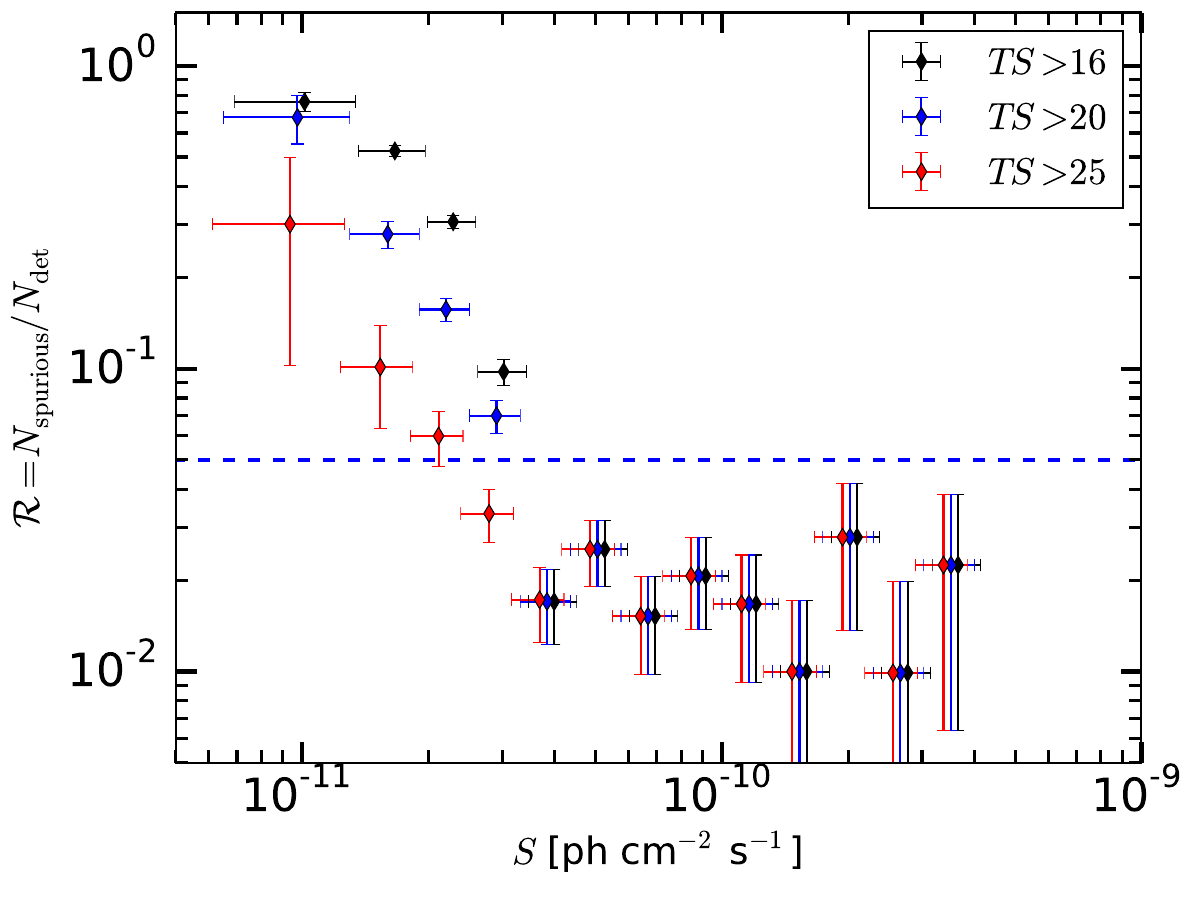}
\caption{Ratio between the number of spurious sources $N_{\rm{spurious}}$ and total number of detected sources $N_{\rm{det}}$ as a function of $S$ for the sources detected with $TS>16/20/25$ (black/blue/red data points) and $5\%$ level that represents the expected fraction of sources missed due to the choice of the $95\%$ CL for the position error (blue dashed line).}
\label{fig:spurious} 
\end{figure}

\subsection{Detection efficiency}
\label{sec:eff}
Given the list of simulated and detected sources for each {\it efficiency simulation}, we are now able to calculate the detection efficiency.
We bin simulated and detected sources in photon flux and we calculate the ratio between the number of detected sources that are real ($N_{\rm{det},i}$) and of simulated sources ($N_{\rm{sim},i}$) in each $i$-th flux bin ($S_i$):
$\omega (S_i) = N_{\rm{det},i}(S^{\rm{obs}}_i) /N_{\rm{sim},i}$.
%
We consider the {\it observed flux} $S^{\rm{obs}}$ for detected sources thus we label this efficiency as {\it observed efficiency} and we indicate it with $\omega$. On the other hand for the simulated sources we use their true flux with which they are simulated.
We calculate $\omega$ considering altogether five simulations and we select only sources that are not spurious using the method explained in Sec.~\ref{sec:spurious}.
In Fig.~\ref{fig:eff} we display $\omega(S)$ for sources detected with $TS > [16,20,25]$. The observed efficiency
$\omega$ is 1 for $S>4\times 10^{-11}$ ph cm$^{-2}$ s$^{-1}$, where the LAT has full detection efficiency. At $S \approx 3 \times 10^{-11}$ ph cm$^{-2}$ s$^{-1}$, $\omega$ is greater than 1 due to the Eddington bias. Indeed, for fluxes lower than this value (see Fig.~\ref{fig:simdet}) many sources are detected with a flux higher than the true flux and at $3 \times 10^{-11}$ ph cm$^{-2}$ s$^{-1}$ the number of detected sources is greater than the number of simulated sources.
\begin{figure}
	\centering
\includegraphics[width=1.03\columnwidth]{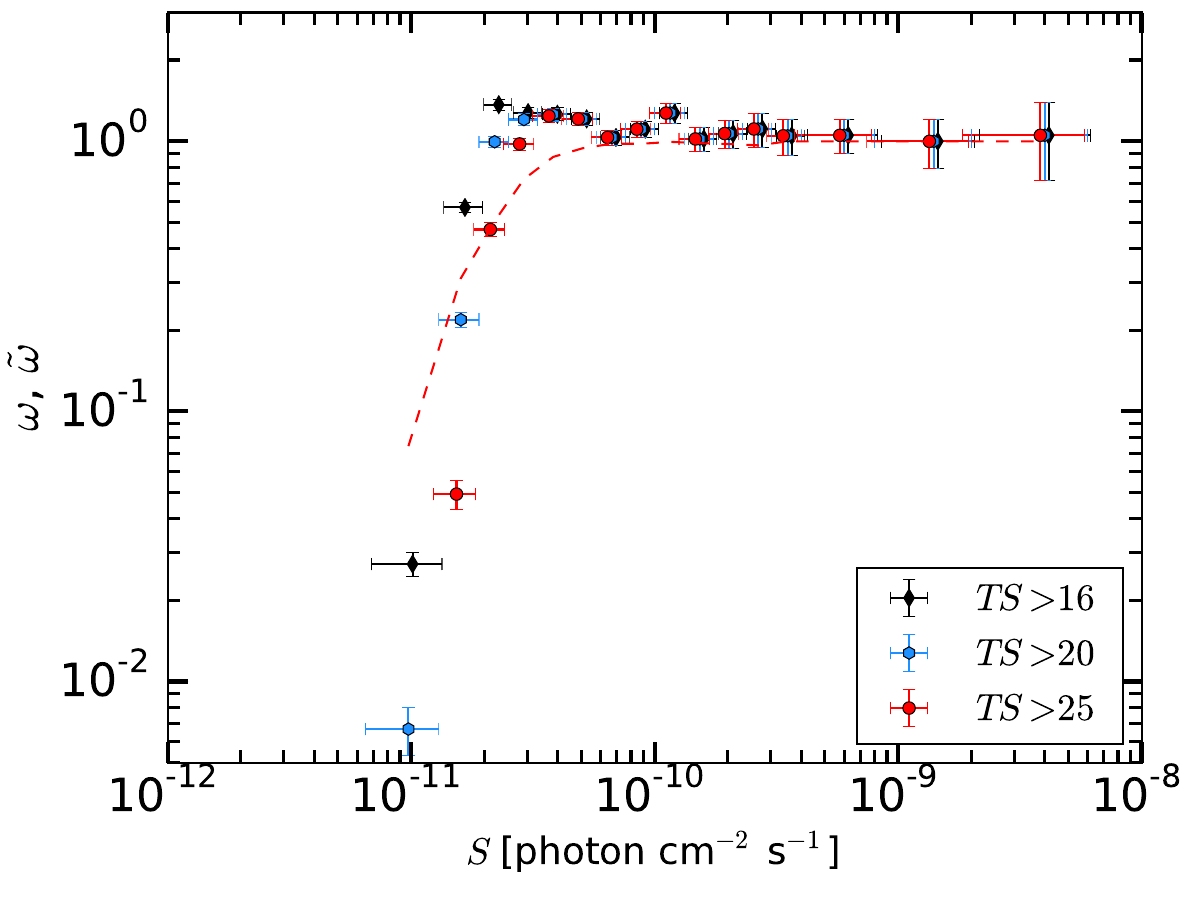}
\caption{{\it Observed efficiency} ($\omega$) for sources detected with $TS>25/20/16$ (red/cyan/black data points). The {\it intrinsic efficiency} $\tilde{\omega}$ calculated with true fluxes is also displayed with red dashed line for sources detected with $TS>25$.}
\label{fig:eff} 
\end{figure}
Below this flux, $\omega(S)$ quickly drops to zero. The detection fraction is, naturally, greater for sources detected with $TS>16$, because of a larger sample containing fainter sources. The results shown in Fig.~\ref{fig:eff} are derived for the Off. IEM. The errors are given by Poisson statistics and are thus statistical.

We also calculate the efficiency considering for the detected sources the flux of the corresponding simulated source ($S^{\rm{true}}$): $\tilde{\omega} (S_i) = N_{\rm{det},i}(S^{\rm{true}}_i)/N_{\rm{sim},i}$. 
We thus derive this quantity in true flux space and we label it as {\it intrinsic efficiency} ($\tilde{\omega}$), shown in Fig.~\ref{fig:eff} for sources detected with $TS>25$. The intrinsic efficiency $\tilde{\omega}$ is always lower or equal to 1, because the Eddington bias is not present in true flux space.

\begin{figure*}[t]
\begin{centering}
 \includegraphics[width=0.49\textwidth]{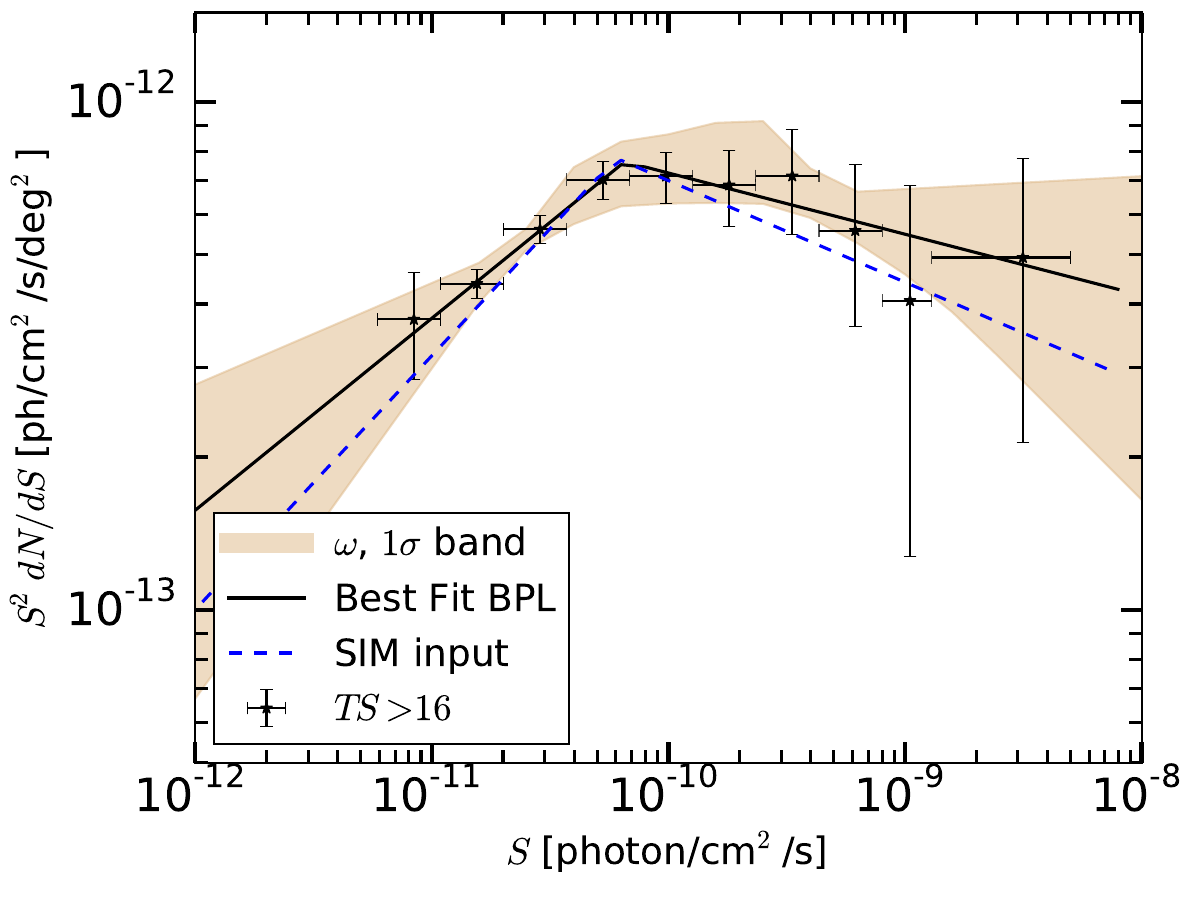}
 \includegraphics[width=0.49\textwidth]{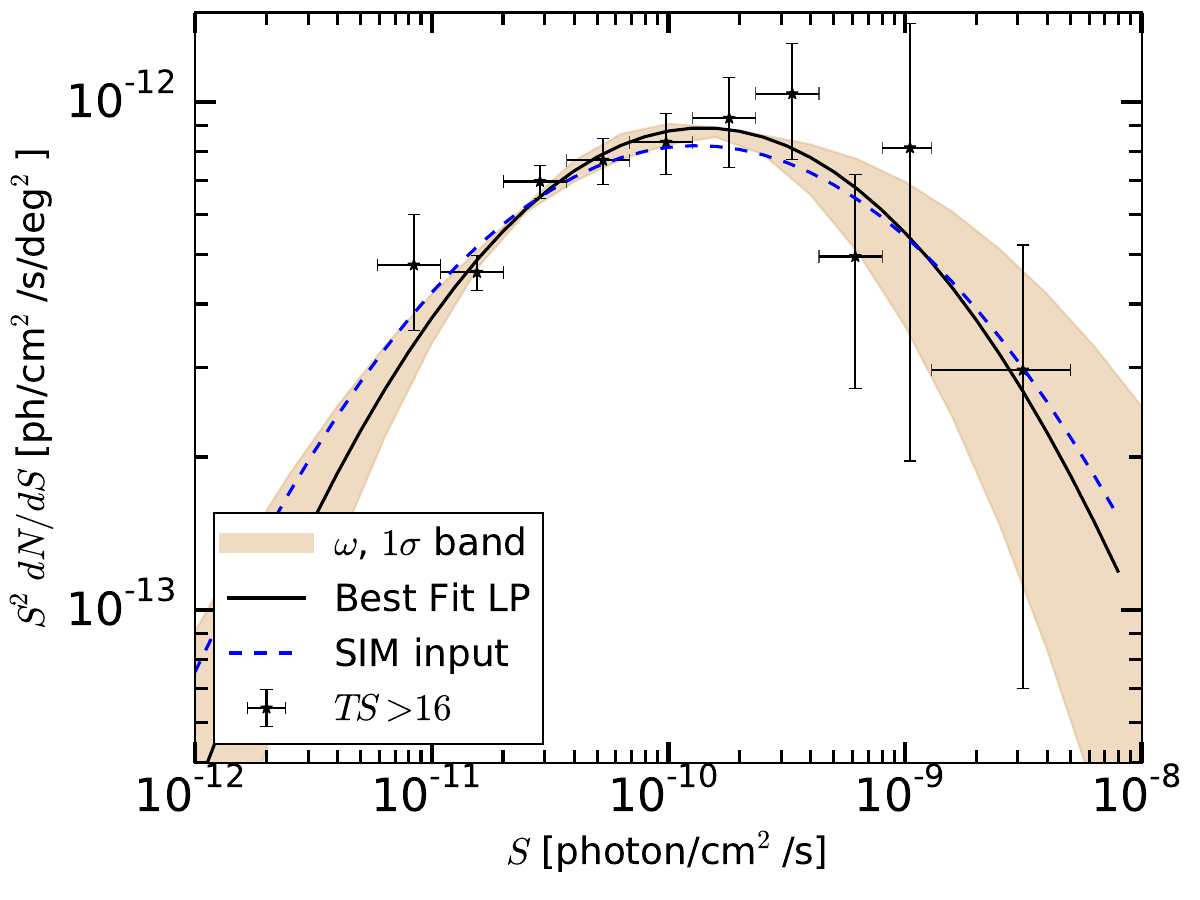}
\caption{Differential source count distribution used in our {\it efficiency validation simulations}. Left (right) panel is for a BPL (LP) $dN/dS$. In each plot we report the input of our {\it efficiency validation simulations} (blue dashed line), the corrected $dN/dS$ (black data), the best fit (black solid line) and $1\sigma$ (brown band) for the fit to the data. 
}
\label{fig:effsim}
\end{centering}
\end{figure*}

\subsection{Validation of the efficiency correction method}
\label{sec:valid}
In this section we use simulations, that we label as {\it efficiency validation simulations}, to validate the efficiency correction method employed to find $\omega$ (see Sec.~\ref{sec:eff}) and the corrected $dN/dS$ (see Sec.~\ref{sec:dNdSeff}).
The {\it efficiency validation simulations} include the Off. IEM and the isotropic template as well as an isotropic distribution of sources with fluxes extracted from a $dN/dS$ distribution parameterized with an BPL or LP shape (see Eq.~\ref{eq:BPL} and \ref{eq:LP}, respectively).
We chose two different shapes in order to check whether we are able to reconstruct both types of $dN/dS$ parameterizations.
We take the flux distribution of sources detected in our {\it efficiency validation simulations} with $TS>16$ and correct it with the {\it observed efficiency} derived in Sec.~\ref{sec:eff}.
Then, we check that the corrected source count distribution is consistent with the input $dN/dS$ shape. 

For the LP, we consider $\gamma=1.90$ and $\beta=-0.10$, while the BPL is given by $\gamma_2=2.20$, $\gamma_1=1.50$, and $S_b =6\times10^{-11}$ ph cm$^{-2}$s$^{-1}$.
The results are shown in Fig.~\ref{fig:effsim} 
where we report the intrinsic $dN/dS$ and the result of the efficiency correction method. We refit the corrected $dN/dS$, finding the LP parameters as $\gamma=1.87\pm0.19$ and $\beta=-0.12\pm0.06$, and the BPL given by $\gamma_2=2.12\pm0.15$, $\gamma_1=1.55\pm0.08$, and $S_b =(6.5 \pm 2.0)\times10^{-11}$ ph cm$^{-2}$s$^{-1}$. The values of the $dN/dS$ parameters are thus compatible within 1$\sigma$ with the input parameters, meaning that the {\it observed efficiency} derived in Sec.~\ref{sec:eff} is well suited for reconstructing a given $dN/dS$. 

\subsection{Study of possible systematics in the {\it observed efficiency}}
\label{sec:effsys}

\begin{figure}
\centering
\includegraphics[width=1.03\columnwidth]{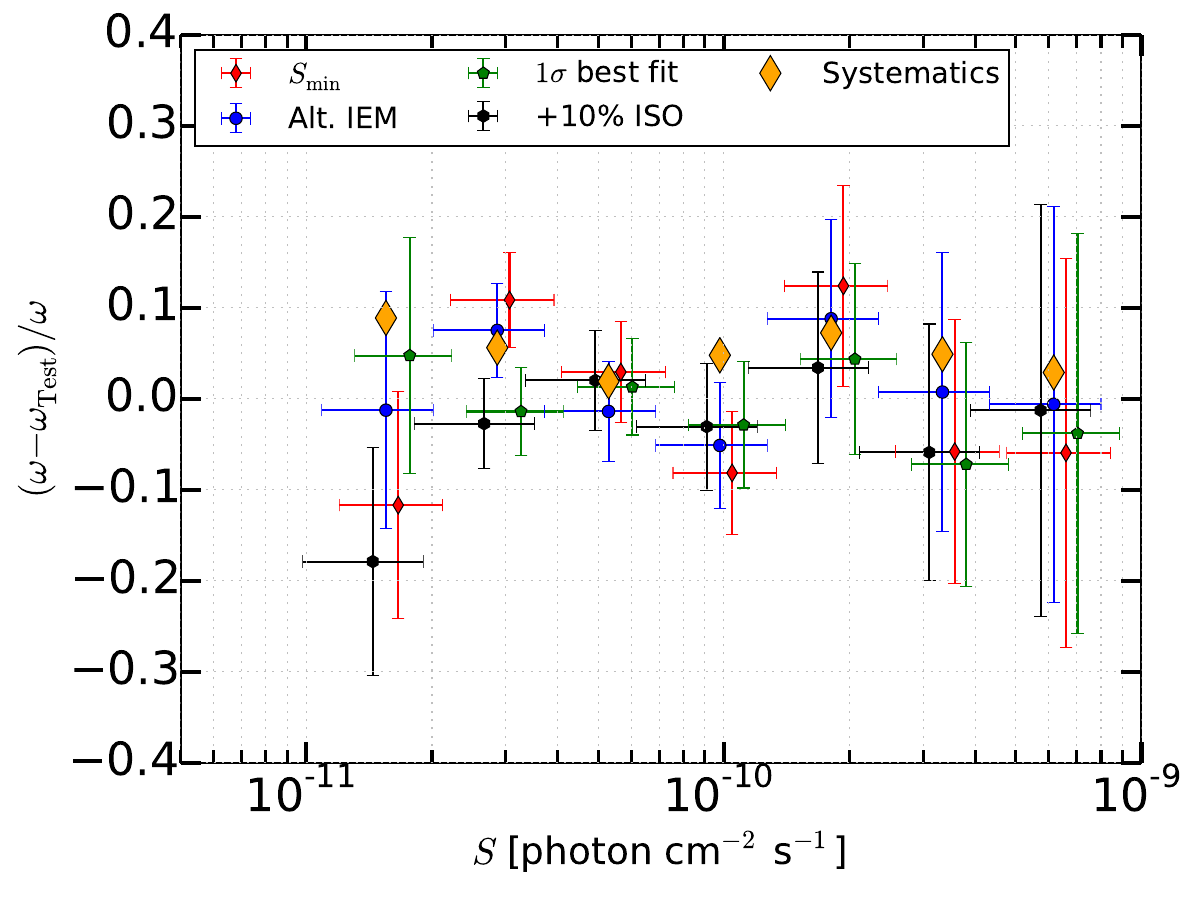}
\includegraphics[width=1.03\columnwidth]{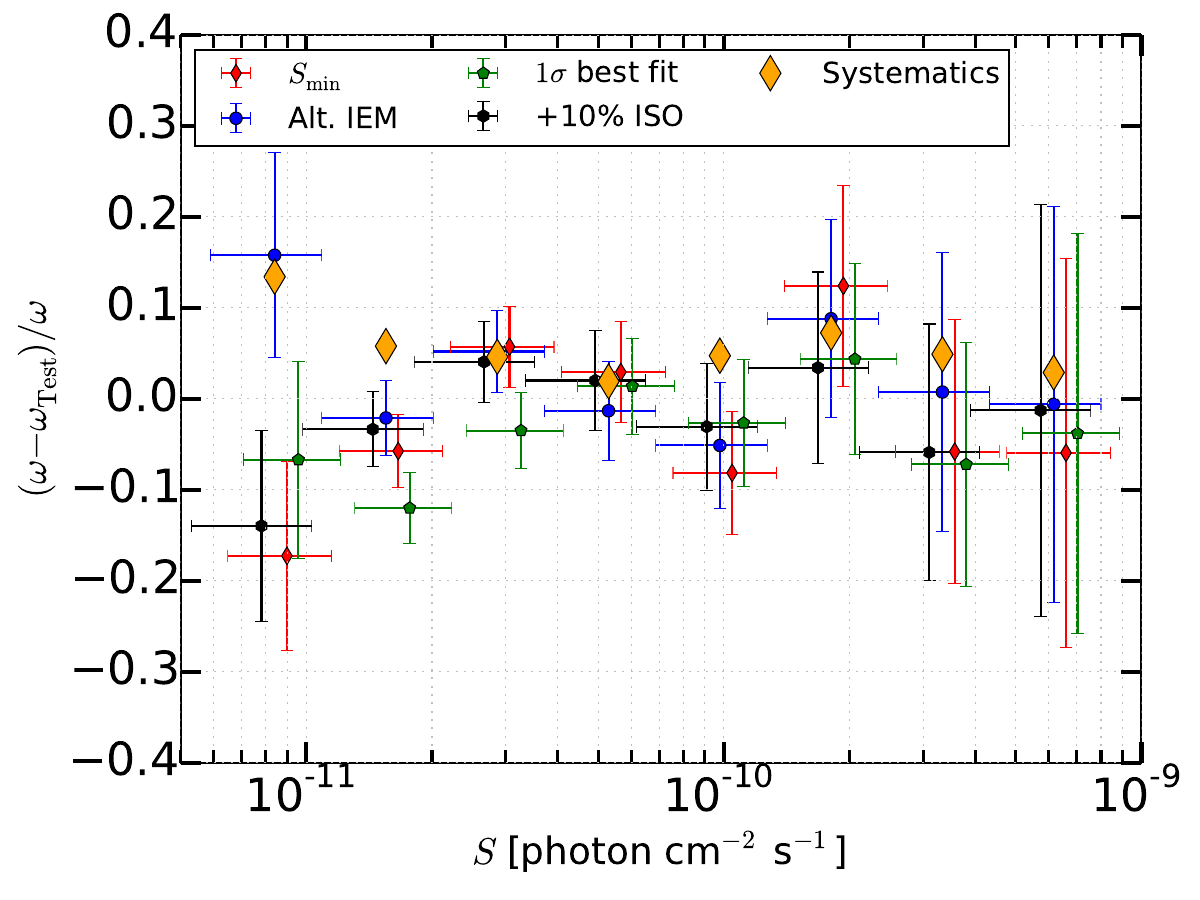}
\caption{Fractional difference $(\omega-\omega_{\rm{Test}})/ \omega$ between the efficiency found with the Off. IEM, $\omega$, and the efficiency calculated with the different test cases explained in the text (see Ref.~\ref{sec:effsys}), $\omega_{\rm{Test}}$. The top (bottom) panel depicts sources detected with $TS>25$ ($TS>16$). The average of the absolute values of $(\omega-\omega_{\rm{Test}})/ \omega$ for the different cases are displayed with yellow diamonds.}
\label{fig:effcheck} 
\end{figure}

We investigate the presence of systematic uncertainties due to the choices that we made in our analysis.
In each of the following scenarios, we generate, analyze, and derive the {\it observed efficiency} sets of five simulations  that we label as {\it systematics simulations}:
\begin{itemize} 
\item \textit{Different IEM.} Simulating the $\gamma$-ray sky with the Alt. IEM and correcting the $dN/dS$ of sources found using the Alt IEM. This run investigates systematics associated to the choice of the IEM.
\item \textit{Different normalization of the isotropic template.} Considering a normalization of the isotropic template that is 10\% greater with respect to the benchmark case. The normalization that we calculate in Sec.~\ref{sec:enspectrum} for the isotropic template has an uncertainty of about $5\%$. Therefore, we test whether having an emission from the isotropic component, that is the component most degenerate with undetected sources, $2\sigma$ greater than the best fit value affects our results.
\item \textit{Lower $S_{\rm{min}}$.} Using a value of $S_{\rm{min}}= 5\times 10^{-13}$ ph cm$^{-2}$ s$^{-1}$. In Sec.~\ref{sec:sim} we draw source fluxes with $S_{\rm{min}}= 7\times 10^{-13}$ ph cm$^{-2}$ s$^{-1}$ and we check if simulating sources with lower fluxes changes our results.
\item \textit{Different $dN/dS$ simulated.} Simulating sources with $\gamma_2=2.15$, $\gamma_1=1.40$ and $S_b=5\times10^{-11}$ ph cm$^{-2}$ s$^{-1}$. This $dN/dS$ is a variation by $1\sigma$ from the best fit of the source count distribution we find in the next section. With this test we check how much the results depend on the shape of the simulated $dN/dS$.
\end{itemize}

The procedure that we use to derive $\omega$ for the above-mentioned test cases is exactly the same that we employ for our benchmark case with the Off. IEM.
Fig.~\ref{fig:effcheck} shows the fractional efficiency difference between the test cases ($\omega_{\rm{Test}}$) and the benchmark case derived with the Off. IEM ($\omega$), for sources detected with $TS>16$ and $TS>25$: $(\omega-\omega_{\rm{Test}})/\omega$. 
The error bars depict statistical errors while the deviation from $(\omega-\omega_{\rm{Test}})/ \omega=0$ corresponds to the systematic error with respect to our benchmark case.
$(\omega-\omega_{\rm{Test}})/ \omega$ is always compatible with 0 within the statistical uncertainties, meaning that there are no clear deviations from the benchmark efficiency except for the lowest flux bin where the differences are of the order of $17\%$ for $TS>16$ and $10\%$ for $TS>25$. These percentages can be considered as systematic uncertainties of our efficiency and are calculated as the average of the absolute values of $(\omega-\omega_{\rm{Test}})/ \omega$ among the four test cases.

\subsection{$dN/dS$ derived with efficiency corrections}
\label{sec:dNdSeff}
We can now correct the source count distribution of our list of detected sources for the {\it observed efficiency} $\omega$, in order to find the real 
source count distribution of the 3FHL:
\begin{equation}
\frac{dN}{dS} \left( S_i \right) = \frac{1}{\Omega \Delta S_i} \frac{N_{i}(1-\mathcal{R}_{i})}{\omega(S_i)} ,
\end{equation}
where $\Omega$ is the solid angle of the $|b|>20^{\circ}$ sky, $\Delta S_i$ is the width of flux bin $i$, $N_{i}$ is the number of sources in the catalog and $\mathcal{R}_i$ is the fraction of spurious sources in each flux bin, and $S_i$ is the flux at the center of the $i$-th bin.
In Fig.~\ref{fig:dNdSsim} we show the corrected source count distribution for both detected sources in the simulations and in the real sky, considering $TS>16$ and $TS>25$. For fluxes above $4\times 10^{-11}$ ph cm$^{-2}$s$^{-1}$, the $dN/dS$ of sources detected in the {\it efficiency simulations} (red points) follows the curve of the source count distribution used as a simulation input. Below this flux, data points drop because the efficiency decreases rapidly and faint sources are difficult to detect.
The second point to notice is that the $dN/dS$ of sources detected from the simulations is perfectly compatible with the input of the {\it efficiency simulation}. This should always be the case since the efficiency is derived from the {\it efficiency simulations}.
Finally, the $dN/dS$ of sources detected from the real sky also follows the shape of the {\it efficiency simulations}, showing that we have simulated a source count distribution that is similar to the real one.

\begin{figure}
	\centering
\includegraphics[width=1.03\columnwidth]{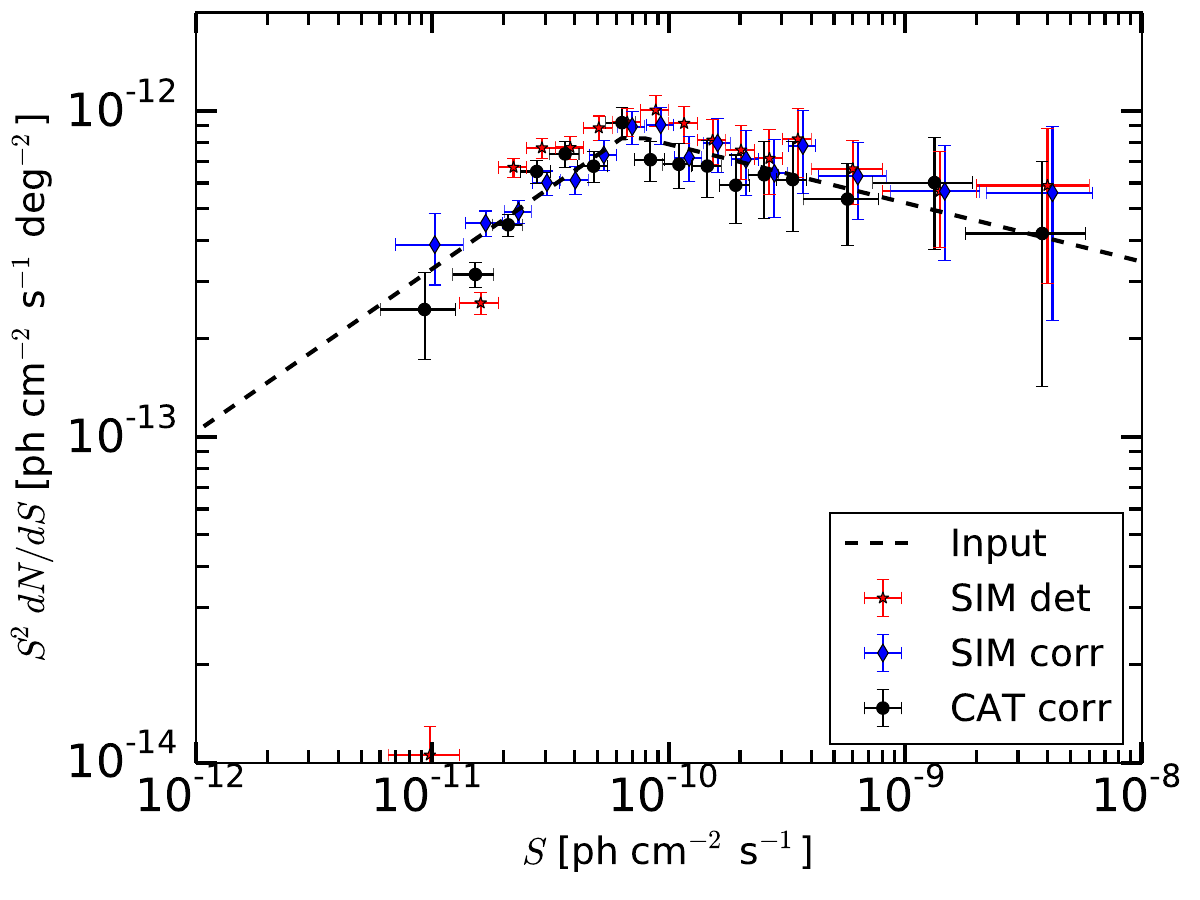}
\includegraphics[width=1.03\columnwidth]{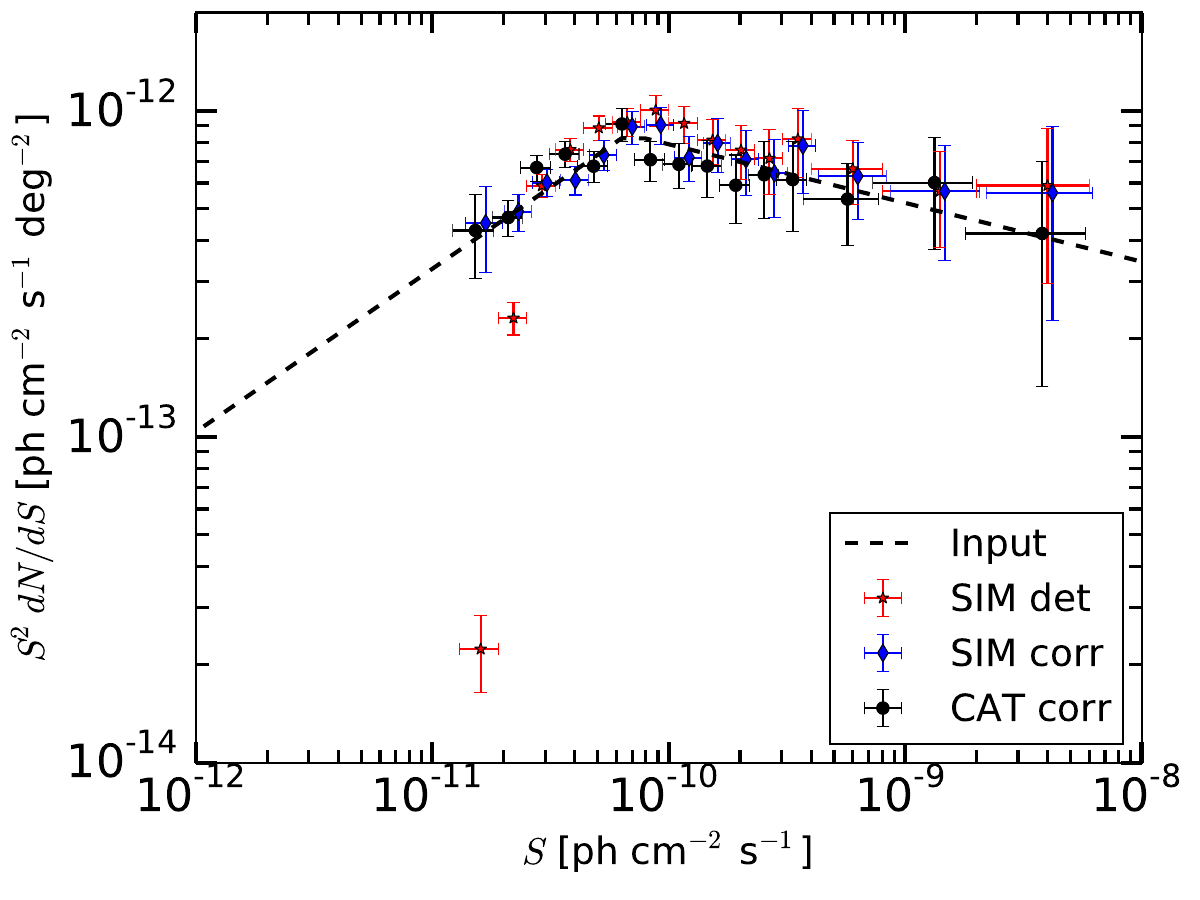}
\caption{Source count distribution of sources detected from the {\it efficiency simulations}, compared to $dN/dS$. The data points depict the uncorrected (red stars) and the efficiency-corrected (blue diamonds) $dN/dS$ of the simulations. The $dN/dS$ of the sources detected in the real data is shown by the black circles. The shape of the simulated $dN/dS$ is shown by the black dashed line. The top (bottom) panel is for $TS>16$ ($TS>25$).}
\label{fig:dNdSsim} 
\end{figure}

We fit $dN/dS$ using different parameterizations. First, we try to fit the data with a PL that gives a chi-square of $\chi^2=21/34/69$ with $13/14/14$ degrees of freedom (d.o.f.) and for $TS>25/20/16$. This translates into a $p$-value of $3\times10^{-9}$ ($2\times10^{-3}$) for the case of $TS>16$ ($TS>20$) and therefore strongly disfavors this scenario. A BPL shape fits the $dN/dS$ well, giving $\chi^2=2.4/3.7/3.1$ 
with $12/13/13$ d.o.f. and for $TS>25/20/16$. In Tab.~\ref{tab:dNdSfit} we summarize the best-fit values of the indexes and the flux breaks. The values of $\gamma_2$, $\gamma_1$, and $S_b$ are consistent among each other for the three datasets.
The flux break is at $\sim3.5\times 10^{-11}$ ph cm$^{-2}$s$^{-1}$ and the slopes above and below the break are about 2.09 and 1.07 for the sample of sources detected at $TS>20$.
Moreover, we calculate the significance for the presence of the break by taking the difference between the $\chi^2$ values for the PL and BPL parameterizations and by comparing that $\Delta \chi^2$ to the two additional d.o.f.~of the BPL shape. This procedure gives $4.2/5.4/7.9 \sigma$ for $TS>25/20/16$. The presence of a flux break is thus significant.
We also test an LP parameterization for the $dN/dS$ as given in Eq.~\ref{eq:LP}, finding it to be slightly disfavored with respect to the BPL ($\Delta \chi^2=10$ for $TS>20$).
The value of the $\chi^2$ for the fit with a BPL is much smaller than the number of degrees of freedom, meaning that the uncertainties on the observed $dN/dS$ are overestimated or that correlations are present between flux bins.
Reducing the magnitude of the uncertainties or taking into account the bin-to-bin correlations would have the result of increasing the $\chi^2$ values and the significance of the break. The results that we present for the significance of the break are thus conservative.

In Fig.~\ref{fig:dNdSfit}, left panel, we show the $dN/dS$ data sets for the three different $TS$ cuts, together with the BPL fit for the case of the $TS>16$ data. The $1\sigma$ statistical uncertainty band of the BPL fit is also displayed. Comparing the results derived for $TS>25/20/16$, we notice that the corrected $dN/dS$ is compatible among the three cases, and that $TS>16$ significantly improves the precision of the derived source count distribution. Likewise, with $TS>16$ we reach a sensitivity of $7.5\times 10^{-12}$ ph cm$^{-2}$s$^{-1}$, allowing us to determine the shape of $dN/dS$ over almost three orders of magnitude.

\begin{figure*}[t]
\begin{centering}
 \includegraphics[width=0.49\textwidth]{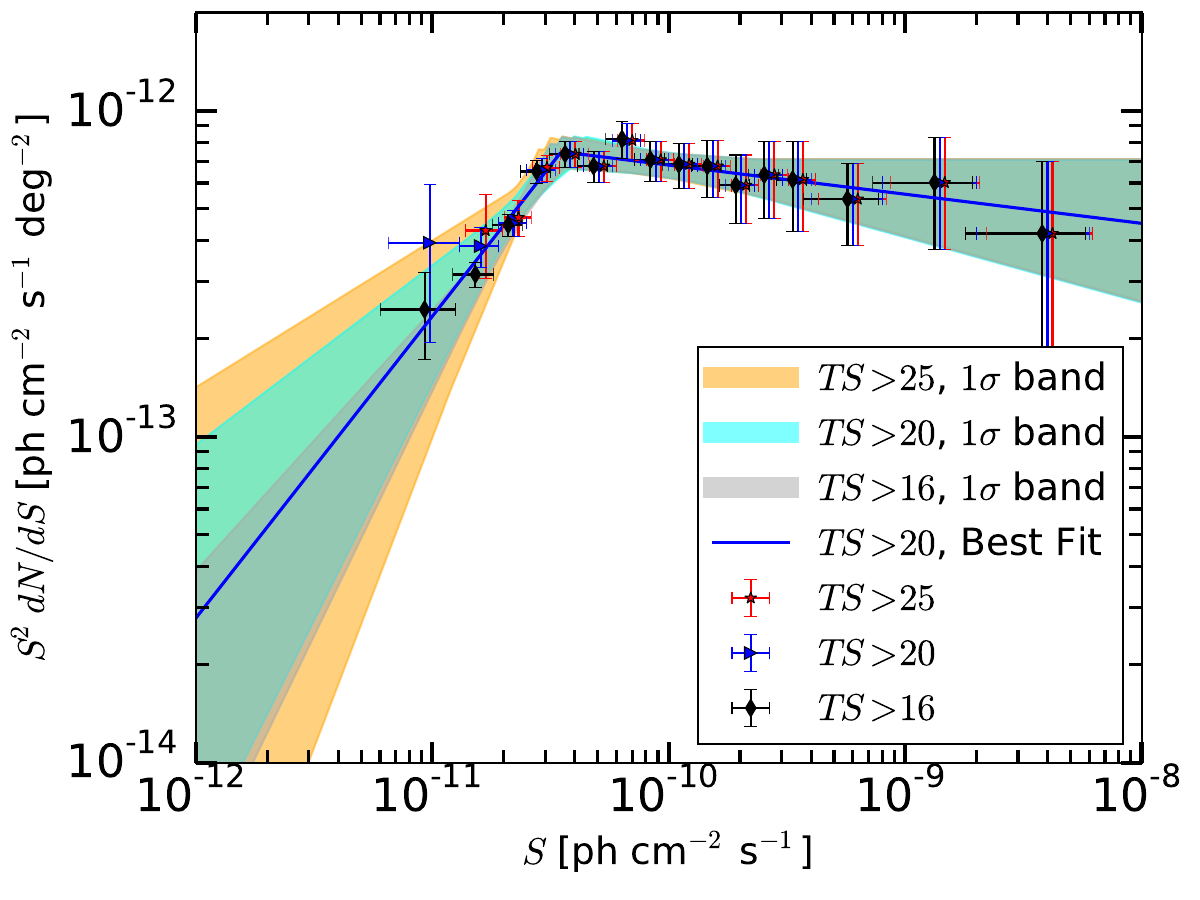}
 \includegraphics[width=0.49\textwidth]{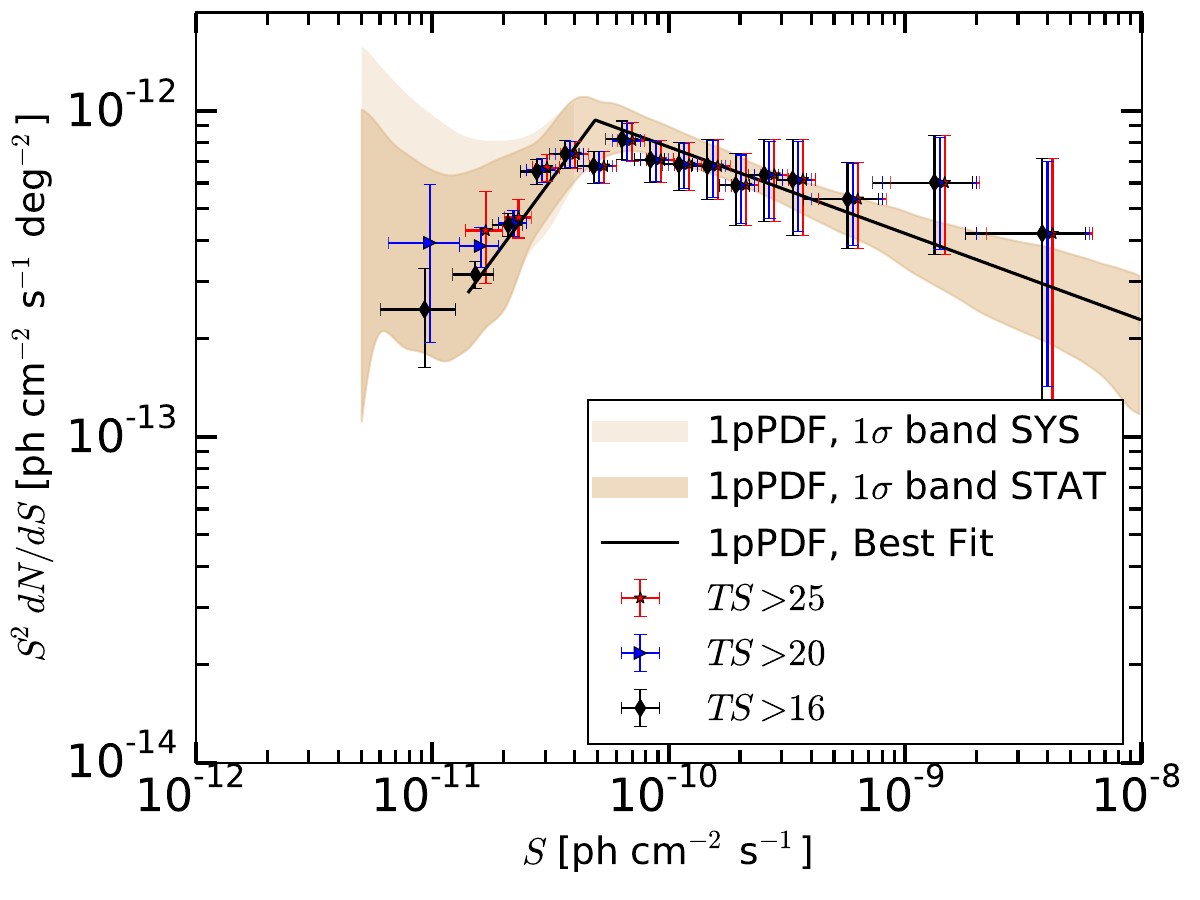}
\caption{Differential source-count distribution $dN/dS$ obtained from 7-year \textit{Fermi}-LAT data between 10 GeV and 1 TeV. Left panel: the result obtained with the efficiency correction method is displayed. The $dN/dS$ distributions as found for $TS>16, 20, 25$ are displayed with black diamonds, blue triangles, and red stars, respectively. The best fit (blue solid line) for the case $TS>20$ and the $1\sigma$ statistical uncertainty band for the cases $TS>16, 20, 25$ are also reported. Right panel: same as in the left panel, but showing the best fit (solid black line) and $1\sigma$ statistical uncertainty band (brown band) for the result obtained with the 1pPDF method. The additional light orange band depicts an estimate of systematic uncertainties using different IEMs.}
\label{fig:dNdSfit}
\end{centering}
\end{figure*}

\begin{figure}[t]
\begin{centering}
\includegraphics[width=1.03\columnwidth]{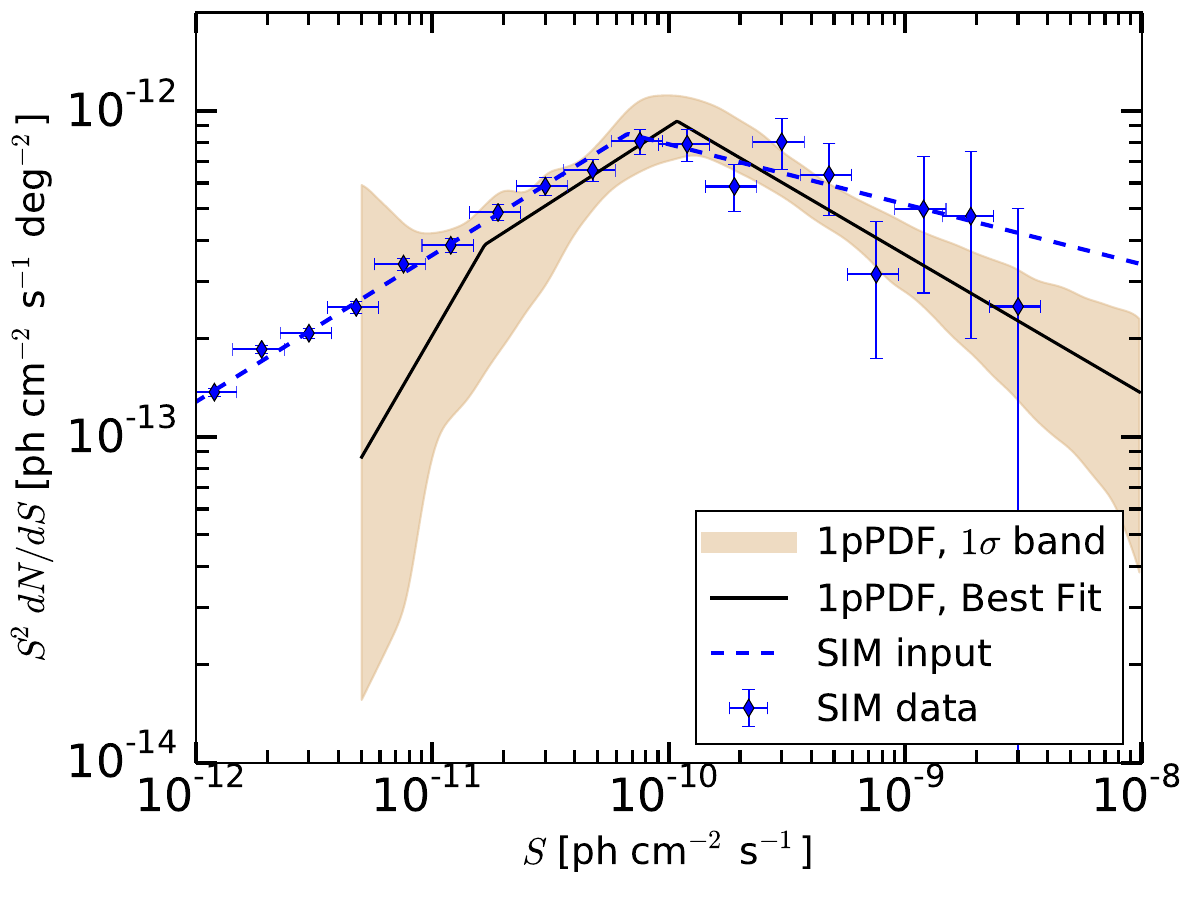}
\caption{1pPDF method applied to one of the simulated data sets discussed in Section~\ref{sec:sim}. The input $dN/dS$ distribution is given by the dashed blue line, while the realization of simulated point sources is depicted by the blue diamonds. The 1pPDF best fit is shown with the solid black line, the orange band shows the corresponding statistical uncertainty at 68\% confidence level.}
\label{fig:dNdS1pPDFsim}
\end{centering}
\end{figure}

\begin{table*}
\center
\begin{tabular}{cccccc}
$TS$ & $\gamma_1$        & $\gamma_2$ & $S_{b}$\,[ph cm$^{-2}$s$^{-1}$]        & $\sigma_{b}$  &  $\mathcal{I}$/EGB (\%)     \\
\hline
$>25$	&	$0.96\pm0.40$ &	$2.09\pm0.04$ &  $(3.3\pm0.4)\cdot10^{-11}$ & 4.2 &$40\pm9$\\
$>20$	&	$1.07\pm0.27$ &	$2.09\pm0.04$ &  $(3.5\pm0.4)\cdot10^{-11}$ & 5.4 &$42\pm8$  \\
$>16$	&	$0.86\pm0.19$ &	$2.09\pm0.04$ &  $(3.4\pm0.3)\cdot10^{-11}$ & 7.9 &$44\pm8$  \\
\end{tabular}
\caption{Fit results for the $dN/dS$ distribution as found with the efficiency correction method. Each row represents the result of the $dN/dS$ fit for different $TS$ cuts. The columns list the best fit and statistical $1\sigma$ errors for the index above ($\gamma_2$) and below ($\gamma_1$) the flux break $S_b$. The last columns list the significance of the flux break in Gaussian sigma and the fractional flux that point sources contribute to the EGB.}
\label{tab:dNdSfit}
\end{table*}

We are now able to calculate the contribution of blazars to the EGB. 
In order to do so we sum the flux of sources in our catalogs ($\mathcal{F}_{\rm{PS}}$, i.e. the emission from detected sources) to the flux of unresolved sources that we calculate integrating the corrected source count distribution and $1-\omega$:
\begin{equation}
\mathcal{I}= \mathcal{F}_{\rm{PS}} + \int_{S_{\rm{min}}}^{S_{\rm{max}}} S \frac{dN}{dS} (1-\omega) dS,
\end{equation}
where $S_{\rm{max}}$ is the flux of the brightest detected source, $S_{\rm{max}}=6\times10^{-9}$ ph cm$^{-2}$s$^{-1}$, and $S_{\rm{min}}$ is the sensitivity of the efficiency correction method given by the flux of the faintest source,
namely $S_{\rm{min}}=7.5\times 10^{-12}$ ph cm$^{-2}$s$^{-1}$ for $TS>16$.
For our sample of sources detected with $TS>16/20/25$, the total contribution to the EGB ($\mathcal{I}$) is $(1.25\pm0.25)/(1.18\pm0.22)/(1.13\pm0.22)\times 10^{-8}$ ph cm$^{-2}$s$^{-1}$sr$^{-1}$, corresponding to $44/42/40\%$ of the EGB (see last column of Tab.~\ref{tab:dNdSfit}).
We remark that we have not used any extrapolation below the sensitivity of the efficiency correction method.

\section{1pPDF Method}\label{sec:1ppdf}
The intrinsic statistical properties of the $\gamma$-ray emission detected in the sky provide a complementary observable of its composition. As a second analysis, we therefore consider the statistical 1pPDF method \citep{Zechlin:2015wdz,Zechlin:2016pme} to measure the source-count distribution $dN/dS$. The 1pPDF represents the statistical distribution of photon counts in individual pixels of the counts map measured with the LAT. The $\gamma$-ray sky is given by a superposition of the same three components as considered for the efficiency correction method, i.e. a population of point sources $dN/dS$, the IEM, and a component of diffuse isotropic emission. Each of the three components imprints on the pixel photon counts and thus on the global statistical map properties in a unique way, and can therefore be detected by means that are complementary to the efficiency correction method.

\subsection{Setup}\label{ssec:1ppdf_setup}
We follow the approach of \citet{Zechlin:2015wdz,Zechlin:2016pme} to perform the 1pPDF analysis of our data, using similar assumptions. Here, we present results considering the most conservative data selection cuts, i.e. UCV event selection along with the PSF3 quartile (see Section~\ref{sec:datasel}). The counts map is pixelized using HEALPix with resolution order~7. Results for the two other event samples and order 8 pixelization are discussed in the Appendix. The $dN/dS$ distribution is parameterized with a multiply broken power law \citep[MBPL; see Eq.~10 in][]{Zechlin:2015wdz}, keeping its parameters free to vary. For the IEM, we compare the two templates discussed in Section~\ref{sec:datasel} (Off. and Alt.). The overall normalization of the IEM, $A_\mathrm{iem}$, is treated as free fit parameter, in order to account for possible model inconsistencies. The diffuse isotropic background is modeled with a PL of photon index 2.3, with its normalization free to vary.

The MBPL representation of $dN/dS$ is implemented with two free consecutive breaks covering the flux range above $3\times 10^{-12}\,\mathrm{ph}\,\mathrm{cm}^{-2}\mathrm{s}^{-1}$, i.e. the range in which sources have been detected with the efficiency correction method. Likewise, the 1pPDF method is expected to lose its sensitivity for sources with much fainter fluxes. Thus, the flux range below $3\times 10^{-12}\,\mathrm{ph}\,\mathrm{cm}^{-2}\mathrm{s}^{-1}$ is parameterized with two breaks at fixed positions (called nodes), in order to technically stabilize the sampling of the 1pPDF likelihood function. The indexes of the PL components connecting the breaks of the MBPL are considered free to vary. Below the last node at $3\times 10^{-13}\,\mathrm{ph}\,\mathrm{cm}^{-2}\mathrm{s}^{-1}$, the MBPL is represented with a PL component with a fixed index of $-10$, which effectively lets it drop to zero.

\subsection{1pPDF simulation}\label{ssec:1ppdf_sims}
To check the performance of our setup, we first apply the method to simulated data that we label as the {\it 1pPDF simulation}.
The {\it 1pPDF simulation} contains the Off. IEM, the isotropic template, and the emission from point sources with fluxes drawn from a BPL with input parameters $\gamma_2=2.184$, $\gamma_1=1.55$, and $S_\mathrm{b}=6.7\times 10^{-11}\,\mathrm{ph}\,\mathrm{cm}^{-2}\mathrm{s}^{-1}$. 
Given that the 1pPDF by definition measures the distribution of sources contained in the data (as opposed to resolving individual sources), it is sufficient to consider only one simulation and to compare the statistical realization of the simulated $dN/dS$ distribution with the findings of the 1pPDF. The result is shown in Fig.~\ref{fig:dNdS1pPDFsim}. The 1pPDF fit reproduces the simulated $dN/dS$ realization well within uncertainties, including the determination of the model parameters $\gamma_2$, $\gamma_1$, and $S_\mathrm{b}$. The parameter values determined by the fit are $\gamma_2=2.4^{+0.3}_{-0.2}$, $\gamma_1=1.5^{+0.3}_{-0.8}$, and $S_\mathrm{b}=1.1^{+1.4}_ {-0.6}\times 10^{-10}\,\mathrm{ph}\,\mathrm{cm}^{-2}\mathrm{s}^{-1}$.\footnote{The slight bias of $\gamma_2$ can be explained with statistical fluctuations of the simulated $dN/dS$ realization in the bright-source regime.} The best-fit normalization of the IEM template is $A_\mathrm{iem}=1.02 \pm 0.03$, consistent with the input of 1. Below a flux of $\sim2\times 10^{-11}\,\mathrm{ph}\,\mathrm{cm}^{-2}\mathrm{s}^{-1}$, the uncertainty increases markedly and the method loses its sensitivity to constrain the $dN/dS$ distribution.

\subsection{Results for flight data}\label{ssec:1ppdf_results}
The result of the $dN/dS$ fit obtained with the 1pPDF method using the Off. IEM template is depicted by the black line and the orange band in the right panel of Fig.~\ref{fig:dNdSfit}. The line depicts the best-fit $dN/dS$ down to the value of the second auxiliary break, at and below which the fit loses its significance. The orange band shows the statistical uncertainty at 68\% CL, shown for fluxes above $5\times 10^{-12}\,\mathrm{ph}\,\mathrm{cm}^{-2}\mathrm{s}^{-1}$. The uncertainty band widens quickly below the effective sensitivity of the method (see Section~\ref{ssec:1ppdf_sims}) and the fit becomes unconstrained. The efficiency correction method and the 1pPDF method agree well within statistical uncertainty, as demonstrated by the data points included in the figure. The 1pPDF method resolves a single break at $5^{+6}_{-3}\times 10^{-11}\,\mathrm{ph}\,\mathrm{cm}^{-2}\mathrm{s}^{-1}$ with PL indexes $2.3^{+0.2}_{-0.1}$ and $1.0^{+0.9}_{-0.5}$ above and below the break. The numbers match the findings of the efficiency correction method within uncertainties. The normalization factor of the IEM is obtained as $A_\mathrm{iem}=1.13 \pm 0.02$.

As found with the {\it 1pPDF simulation} in Section~\ref{ssec:1ppdf_sims}, the 1pPDF method loses its sensitivity for resolving point sources below fluxes of $\sim2\times 10^{-11}\,\mathrm{ph}\,\mathrm{cm}^{-2}\mathrm{s}^{-1}$. The analysis of the flight data supplements this sensitivity estimate, given that the uncertainty significantly increases below that value. The sensitivity of the efficiency correction method is obtained to be similar or slightly better, as demonstrated by a comparison with the $TS>16$ data. From the perspective of event statistics, the efficiency correction method profits from significantly higher photon statistics with regard to the stringent data cuts used here for the 1pPDF method ({\tt SOURCE} vs. {\tt ULTRACLEANVETO} event selection and PSF3 restriction, see Tab.~\ref{tab:analysis}. The total number of events differs by a factor of almost 8.). While the event selection cuts for the 1pPDF can be relaxed (see Appendix), it should be emphasized that, for both analysis methods, the real sensitivity for flight data may be primarily driven by systematic uncertainties, as investigated in the following section.

\subsection{Systematic uncertainties}\label{ssec:1ppdf_syst}
To estimate possible systematics related to model choices and assumptions, the analysis setup is tested for stability against the following changes:
\begin{itemize}
\item \textit{Different IEMs}. The analysis is conducted choosing a variety of different IEMs. In particular, we use the alternate IEM described in Section~\ref{sec:datasel}. Other IEMs investigated are models A, B, and C as used in \citet{Ackermann:2014usa} to constrain the systematic uncertainties of the IGRB analysis.
\item \textit{Galactic latitude cut}. We test changes of the Galactic latitude cut to $|b|>10^\circ$ and $|b|>30^\circ$.
\item \textit{Fermi Bubbles/Loop I}. The Galactic plane mask of $|b|>20^\circ$ applied in the main analysis is extended to masking the Fermi Bubbles and Galactic Loop I \citep{2009arXiv0912.3478C}.
\end{itemize}
In summary, we find that our results are stable against all mentioned changes above a flux of $S_\mathrm{syst} \simeq 2\times 10^{-11}\,\mathrm{ph}\,\mathrm{cm}^{-2}\mathrm{s}^{-1}$. Below $S_\mathrm{syst}$, systematic uncertainties begin to increase, mainly driven by uncertainties of the IEM that are depicted by the light orange band in the right panel of Fig.~\ref{fig:dNdSfit}.

Systematics related to the choice of the pixel size are discussed in the Appendix.

\section{Modeling of the Blazar Population}
\label{sec:lf}
Observations at $E>$10\,GeV have the power to constrain models of the blazar population \citep{DiMauro:2013zfa,Ajello:2015mfa}. These models are typically used to study and understand the properties of the blazar family and its evolution and predict the contribution of blazars to the EGB. They comprise two main ingredients: a luminosity function that describes the evolution of the population and a model of the SED that describes their emission as function of energy.  Derivations of the luminosity functions require a complete sample of blazars with full redshift information, which are typically derived from broadband LAT catalogs \citep[i.e. $>$100\,MeV, such as the 3FGL][]{2015ApJS..218...23A}. Because the emission of distant sources is redshifted, the SED models need to be able to describe the emission of blazars on a wider energy range than that of the observation. In \citet{Ajello:2015mfa} the blazar SED is modeled, in the source frame, as a smoothly joined double power law attenuated by the extragalactic-background light as follows:

\begin{equation}
\frac{dN_{\gamma}}{dE}  =   K\left[ 
\left(\frac{E}{E_b} \right)^{\gamma_{a}} +
\left(\frac{E}{E_b} \right)^{\gamma_{b}} \right]^{-1}
\cdot e^{-\tau(E,z)} 
\label{eq:sed}
\end{equation}
In the above work the break energy $E_b$ and low-energy index $\gamma_a$ were chosen to reproduce the observed distributions of curvature parameters\footnote{In LAT catalogs sources that display statistically significant curvature are typically modeled with a log-parabola, whose $\beta$ parameter describes the curvature of the spectrum.} and photon indices of LAT blazars. Among all parameters, the high-energy index $\gamma_b$ that describes the SED at energies even beyond those samples by the LAT, remains the most uncertain. Its original value of $\gamma_b=2.6$ was chosen to reproduce the SED of a few popular blazars (e.g. RBS 0413, Mrk 421, Mrk 501) with contemporaneous GeV--TeV data.  However, contemporaneous observations across a wide wavelength range still only exist of a handful of blazars and are limited, very often, to bright activity states of the sources (which may also be in a spectrally harder state). As such these observations may not be representative of the long-term average SED that blazar population models need.

High-energy catalogs, which are derived using long-term LAT observations, can be useful in constraining the average blazar SED and, in particular, the high energy component. Here we use the luminosity-dependent density evolution (LDDE) and pure luminosity evolution (PLE) models of \citet{Ajello:2015mfa} and change the high-energy index $\gamma_b$ of the SED model until the blazar population model correctly reproduces the observed 3FHL source count distribution. We find that this required an high-energy index $\gamma_b\approx2.8$. Fig.~\ref{fig:dNdSpred} shows the prediction of the \cite{Ajello:2015mfa} evolutive models tuned to reproduce the observed 3FHL source counts.
Both models reproduce the source count distribution obtained with the 1pPDF and efficiency correction methods reasonably well in the flux range of the 1FHL ($S > 5\times 10^{-11} \mathrm{ph}\,\mathrm{cm}^{-2}\mathrm{s}^{-1}$) where the blazar model is tuned \cite{Ajello:2015mfa}.
On the other hand, the source density predicted by the PLE model is a factor of $\approx2$ higher than the one of the PDDE model at fluxes of 10$^{-12}$\,ph cm$^{-2}$ s$^{-1}$.
Moreover, the apparent steeper faint-end slope of the $dN/dS$ obtained with the efficiency correction method may highlight that the blazar population experiences, at least at $>10$ GeV, a stronger evolution than what was found before \cite[see, e.g.,][]{Ajello:2015mfa}.
A $>$10\,GeV catalog that relies on $\sim$20\,yr of LAT exposure or deep surveys with CTA will allow us to discriminate between these two scenarios.

\begin{figure}[t]
\begin{centering}
\includegraphics[width=1.03\columnwidth]{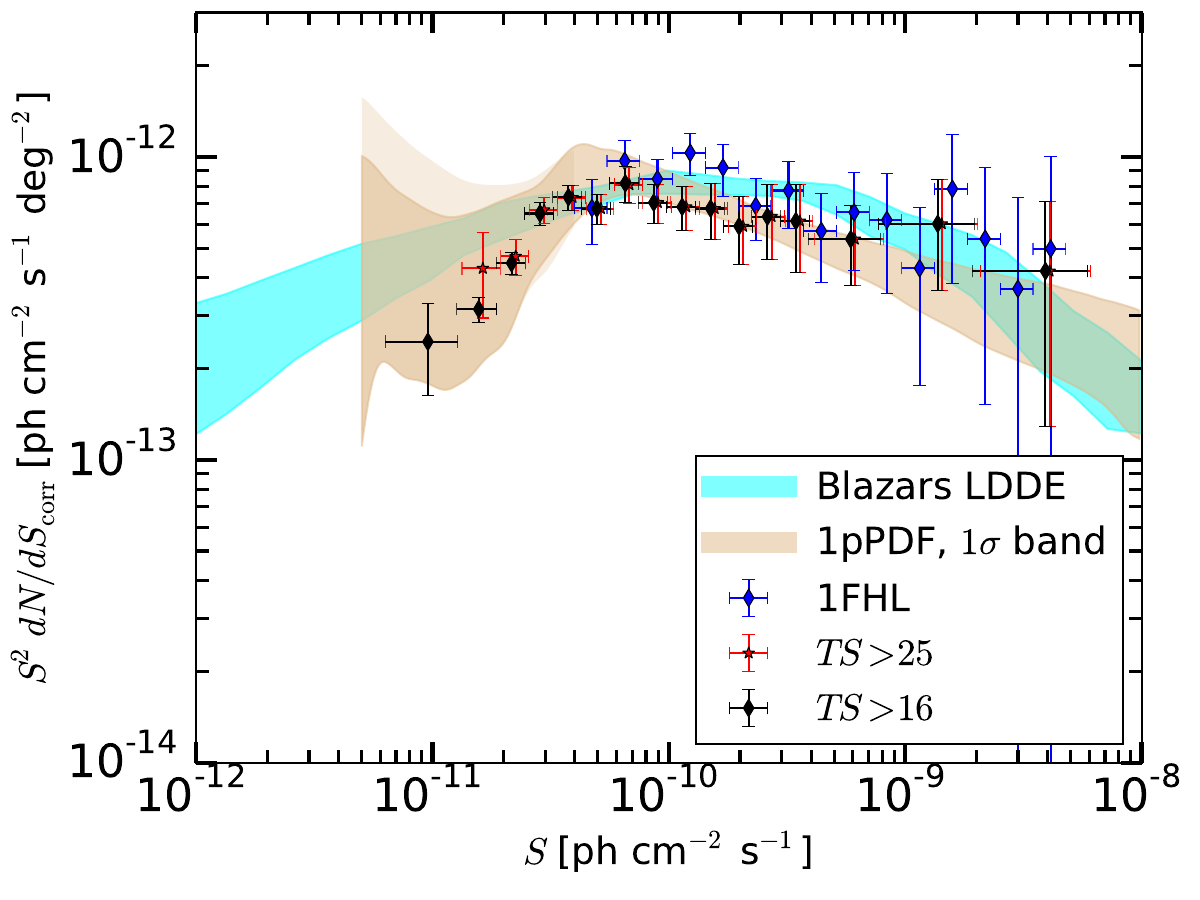}
\includegraphics[width=1.03\columnwidth]{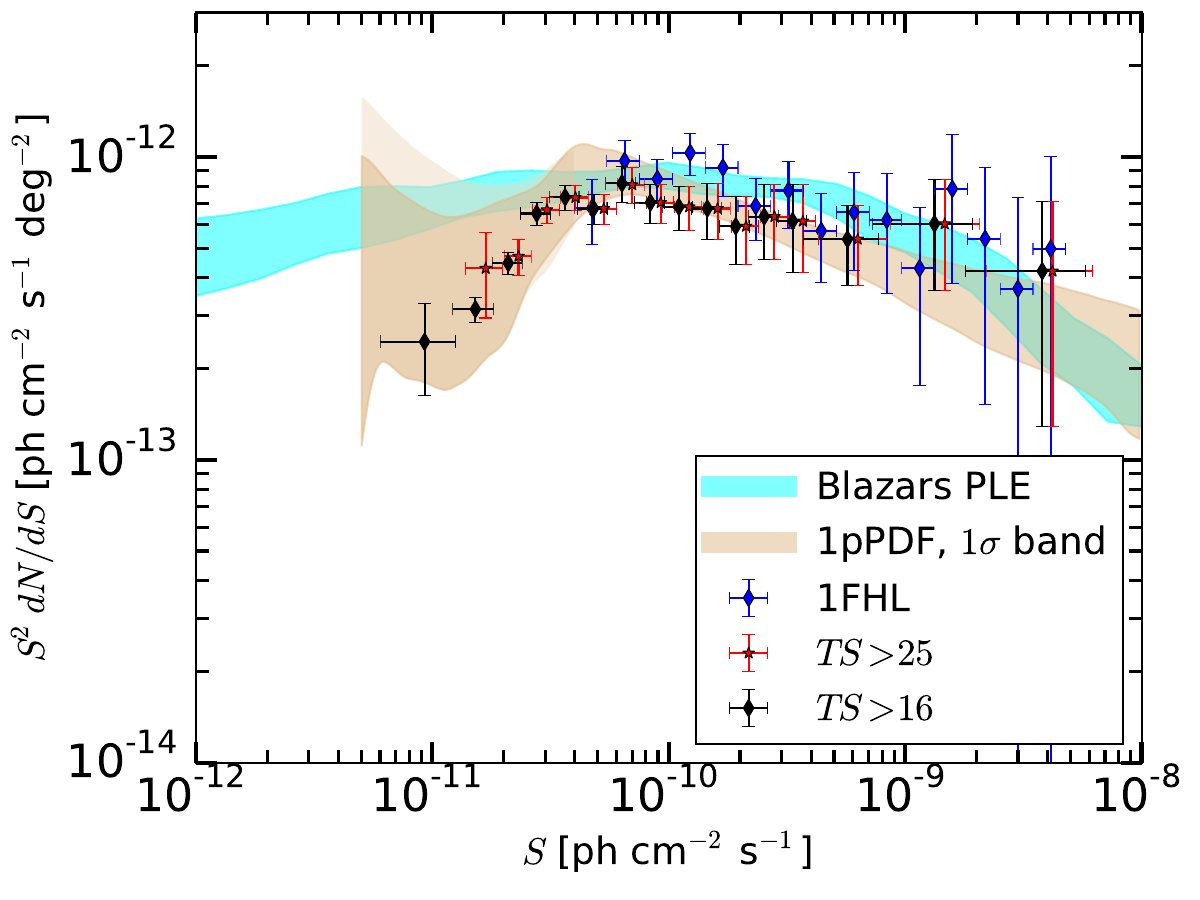}
\caption{Same as in Fig.~\ref{fig:dNdSfit} with the addition of the prediction from blazars with the LDDE (top panel) and PLE (bottom panel). The data derived with efficiency correction method include either statistical and systematic uncertainties.}
\label{fig:dNdSpred}
\end{centering}
\end{figure}

\section{Conclusions}
\label{sec:conclusion}
We have derived the source count distribution $dN/dS$ of the extragalactic $\gamma$-ray sky at $|b|>20^{\circ}$ for energies $E>10$ GeV.
The {\it Fermi}-LAT Collaboration has recently released a new catalog of 1120 sources at $|b|>20^{\circ}$ detected in this energy range, based on 7~years of data, of which most sources are associated with blazars \cite{TheFermi-LAT:2017pvy}.
We have employed efficiency corrections, finding that the $dN/dS$ distribution is well described with a BPL, with a slope above and below a flux break at $(3.5\pm0.4)\times 10^{-11} \mathrm{ph}\,\mathrm{cm}^{-2}\mathrm{s}^{-1}$ of $2.09\pm0.04$ and $1.07\pm0.27$.
These findings have been confirmed with a complementary technique, analyzing the statistics of pixel photon counts with the 1pPDF method. The BPL parameterization of $dN/dS$ is significantly preferred over a single PL parameterization, with a significance for a break of $5.4 \sigma$ (using sources detected with $TS>20$).
The efficiency correction method reaches a sensitivity of $\sim7.5\times10^{-12}\,\mathrm{ph}\,\mathrm{cm}^{-2}\mathrm{s}^{-1}$, while the 1pPDF reaches $10^{-11}\,\mathrm{ph}\,\mathrm{cm}^{-2}\mathrm{s}^{-1}$, permitting to constrain the $dN/dS$ over almost three orders of magnitude in photon flux.
The most direct consequence is that $(42\pm8)\%$ of the EGB emission above 10 GeV and for $S > 7.5\times10^{-12}\,\mathrm{ph}\,\mathrm{cm}^{-2}\mathrm{s}^{-1}$ is composed of resolved and unresolved point sources, which are mainly blazars.
This measurement therefore improves the previous results obtained from the four-year data set \citep{Ackermann:2014usa}, where the contribution of point sources detected by the LAT has been calculated to be $(35 \pm 10)\%$ of the EGB at $E>10$ GeV.

The remaining fraction of the EGB could be attributed to a faint population of sources like misaligned AGN \citep[see, e.g.,][]{DiMauro:2013xta} or star-forming galaxies \citep[see, e.g.,][]{2012ApJ...755..164A}. These two source populations are rare in {\it Fermi}-LAT catalogs but they are expected to include many faint sources that can emerge in the $dN/dS$ as a change in the slope below the current data coverage.
For example, a Euclidean $dN/dS \propto S^{-2.5}$ source count distribution, as expected for misaligned AGN \citep{DiMauro:2013xta}, that leads to a $dN/dS$ break at $2\times 10^{-12}$\,ph cm$^{-2}$ s$^{-1}$, would entirely resolve the EGB for $S>10^{-14}$\,ph cm$^{-2}$ s$^{-1}$.
An alternative interpretation of the remaining flux fraction of the EGB is the purely diffuse $\gamma$-ray emission of ultra high-energy cosmic rays (UHECRs) interacting with the extragalactic background light \citep[e.g.,][]{Gavish:2016tfl}.
The energy spectrum of the contribution from unresolved misaligned AGN and star-forming galaxies is expected to follow the SED of detected sources from this population ($dN/dE \propto E^{-2.4}$).
On the other hand, the spectral shape of the contribution from UHECRs is expected to have a slope smaller than $-2.0$ \citep{Gavish:2016tfl}.
Calculating the contribution of point sources to the EGB in different energy bins will help to single out the correct interpretation and will be addressed by forthcoming work.


\begin{acknowledgements}
  The {\it Fermi} LAT Collaboration acknowledges generous ongoing
  support from a number of agencies and institutes that have supported
  both the development and the operation of the LAT as well as
  scientific data analysis. These include the National Aeronautics and
  Space Administration and the Department of Energy in the United
  States; the Commissariat \`a l'Energie Atomique and the Centre
  National de la Recherche Scientifique/Institut National de Physique
  Nucl\'eaire et de Physique des Particules in France; the Agenzia
  Spaziale Italiana and the Istituto Nazionale di Fisica Nucleare in
  Italy; the Ministry of Education, Culture, Sports, Science and
  Technology (MEXT), High Energy Accelerator Research Organization
  (KEK), and Japan Aerospace Exploration Agency (JAXA) in Japan; and
  the K. A. Wallenberg Foundation, the Swedish Research Council, and
  the Swedish National Space Board in Sweden.  Additional support for
  science analysis during the operations phase is gratefully
  acknowledged from the Istituto Nazionale di Astrofisica in Italy and
  the Centre National d'Etudes Spatiales in France.

  The authors are grateful to Keith Bechtol and Jean Ballet for their insightful comments and suggestions.
  MDM, EC acknowledge support by the NASA {\it Fermi} Guest Investigator
  Program 2014 through the {\it Fermi} multi-year Large Program N. 81303
  (P.I. E.~Charles). MDM, EC and SM acknowledge the NASA {\it Fermi} Guest Investigator
  Program 2016 through the {\it Fermi} one-year Program N. 91245
  (P.I. M.~Di Mauro). SM gratefully acknowledges support by the Academy of Science of Torino through the 
  {\it Angiola Agostinelli Gili} 2016 scholarship and the KIPAC institute at 
  SLAC for the kind hospitality during this project.
  
  HSZ gratefully acknowledges the Istituto Nazionale di Fisica Nucleare (INFN)
  for a post-doctoral fellowship in theoretical physics on
  "Astroparticle, Dark Matter and Neutrino Physics", awarded under the INFN
  Fellowship Programme 2015.
  \end{acknowledgements}

\begin{appendix}
\section{\label{app:1pPDFcuts}1pPDF Method with Relaxed Data Selection Cuts}
\subsection{Event selection}
To supplement the stability of the 1pPDF results obtained in Sections~\ref{ssec:1ppdf_results} and \ref{ssec:1ppdf_syst}, the event sample can be significantly increased by relaxing the data selection cuts (see Section~\ref{sec:datasel}). Figure~\ref{fig:dNdS_1pPDF_xchk} shows the $dN/dS$ distribution measured with the 1pPDF method as applied to the {\tt SOURCE} and {\tt CLEAN} data samples, pixelized with resolution order 7. The 1pPDF analysis configuration matches the setup used for the benchmark results discussed in Section~\ref{ssec:1ppdf_setup}. Due to higher event statistics, for both data samples the statistical uncertainty of $dN/dS$ narrows in the regime down to $\sim 2\times 10^{-11}\,\mathrm{cm}^{-2}\mathrm{s}^{-1}$. An upturn at the very faint end of the distribution seems to appear for {\tt SOURCE} events, though with a high statistical uncertainty. No statistical evidence has been found for this feature. As indicated by the systematic uncertainty (light orange band) obtained with different IEMs, we attribute this feature to most likely originate from un- or mismodeled Galactic foreground fluctuations (at the spatial scale of point sources), demonstrating that the effective sensitivity of this analysis is $\sim 10^{-11}\,\mathrm{cm}^{-2}\mathrm{s}^{-1}$. Nevertheless, the possibility of a physical origin of this feature remains.

\begin{figure*}[t]
\begin{centering}
 \includegraphics[width=0.49\textwidth]{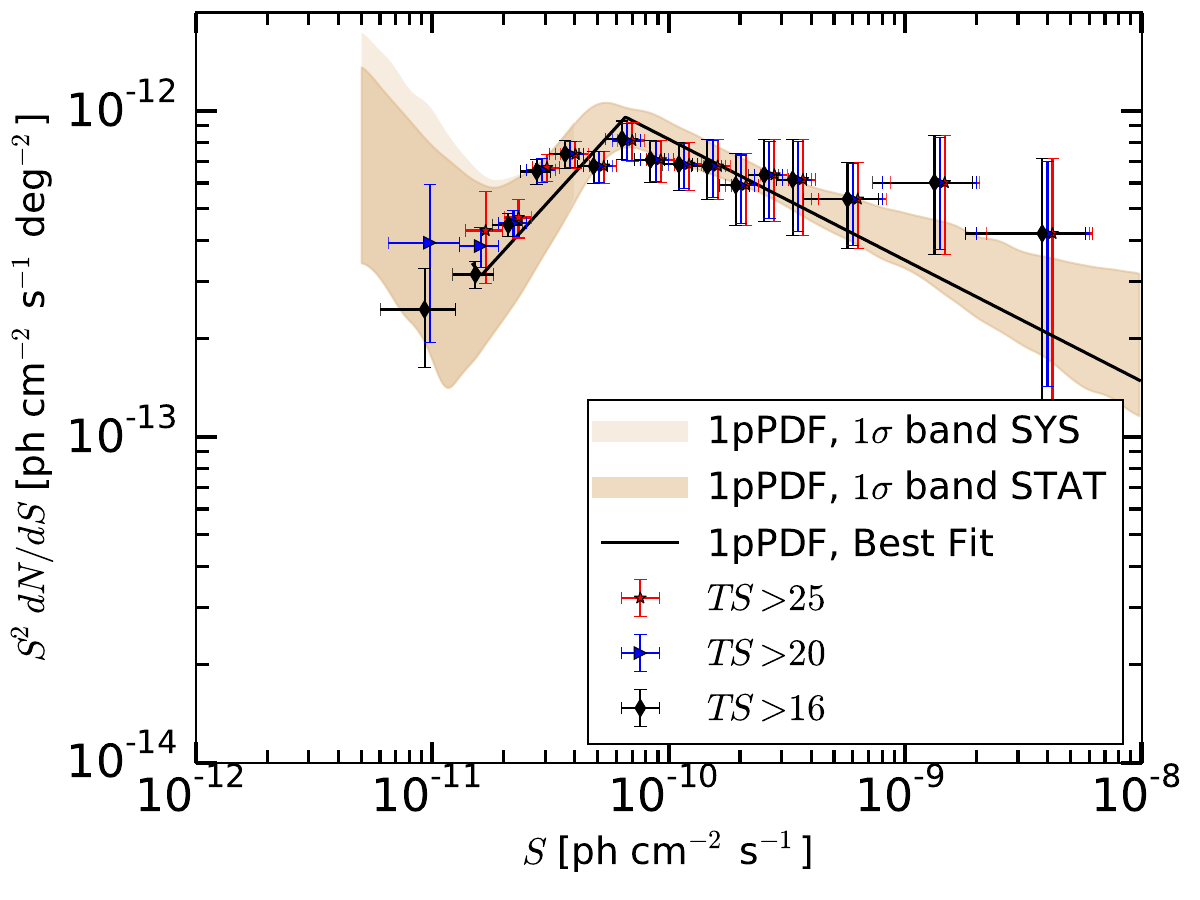}
 \includegraphics[width=0.49\textwidth]{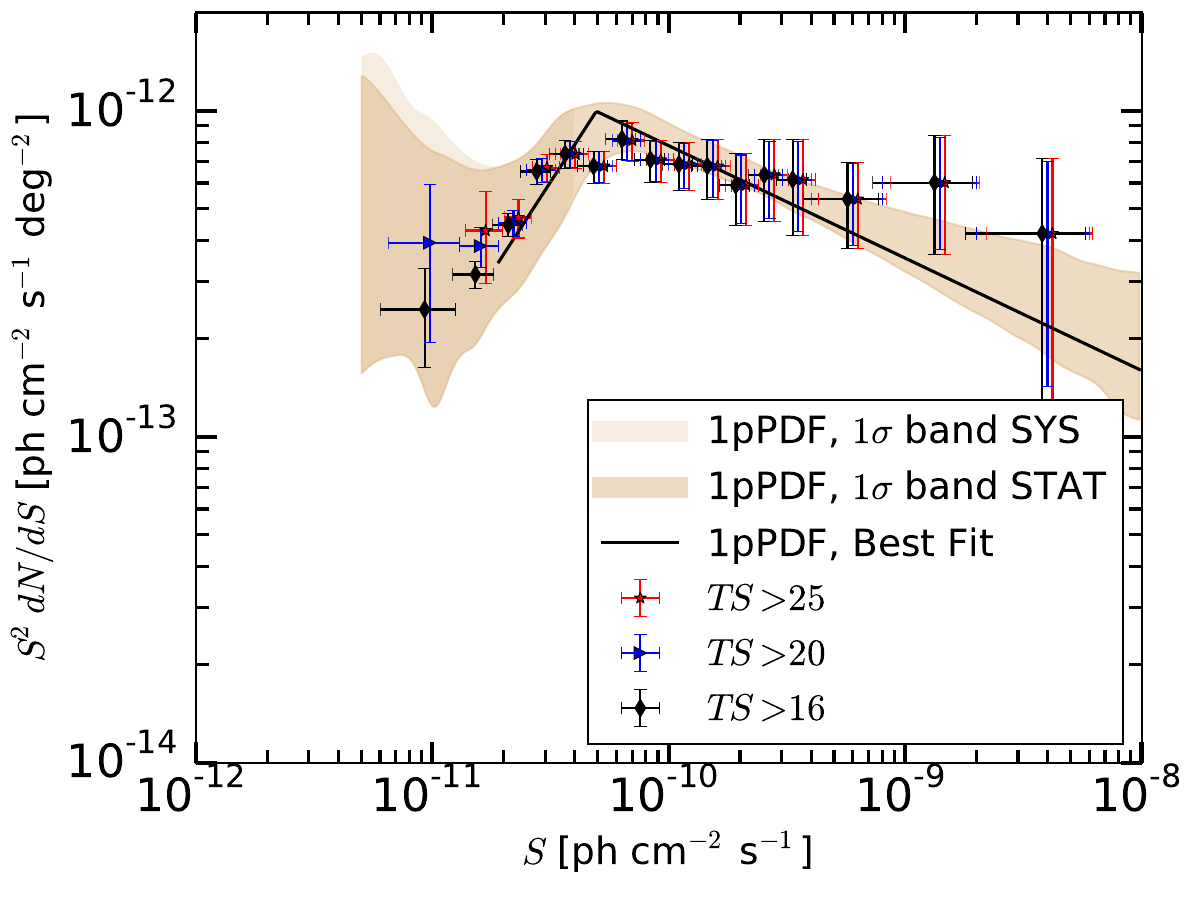}
\caption{Differential source-count distribution $dN/dS$ obtained with the 1pPDF method, using 7-year \textit{Fermi}-LAT data with {\tt SOURCE} (left) and {\tt CLEAN} (right) event selection cuts. Notation and data points are the same as in the right panel of Figure~\ref{fig:dNdSfit}.}
\label{fig:dNdS_1pPDF_xchk}
\end{centering}
\end{figure*}

\subsection{Pixel size}
As demonstrated by past studies of the 1pPDF method \citep[][]{Zechlin:2015wdz,Zechlin:2016pme}, PSF smoothing can pose a systematic limit to the 1pPDF setup. This effect is caused by the finite PSF of the instrument, distributing detected photons from point-like sources over a certain sky area. The average PSF of a data sample in a given energy bin, which is corrected for by the 1pPDF setup, is influenced by the PSF and the exposure as functions of energy, the width of the energy bin, and the shapes of the point source's emission spectra. Furthermore, the PSF of the LAT depends on other event characteristics such as the event impact angle. The relative broadness of the average sample PSF is to be seen in relation to the used pixel size, and thus, vice versa, an optimum pixel size can be defined for the 1pPDF analysis. Previous studies have shown that the systematics related to PSF smoothing can be reduced by undersampling the effective PSF width (i.e. choosing a lower grid resolution with respect to the effective PSF size). Our choice of resolution order~7 is driven by these results. In addition, Figure~\ref{fig:1pPDF_hp8} depicts results obtained for the same benchmark UCV data as in Section~\ref{sec:1ppdf}, but using pixel resolution order~8. The sensitivity of this analysis slightly decreases, given the larger uncertainty band below the break.

\begin{figure}[t]
\begin{centering}
\includegraphics[width=0.49\columnwidth]{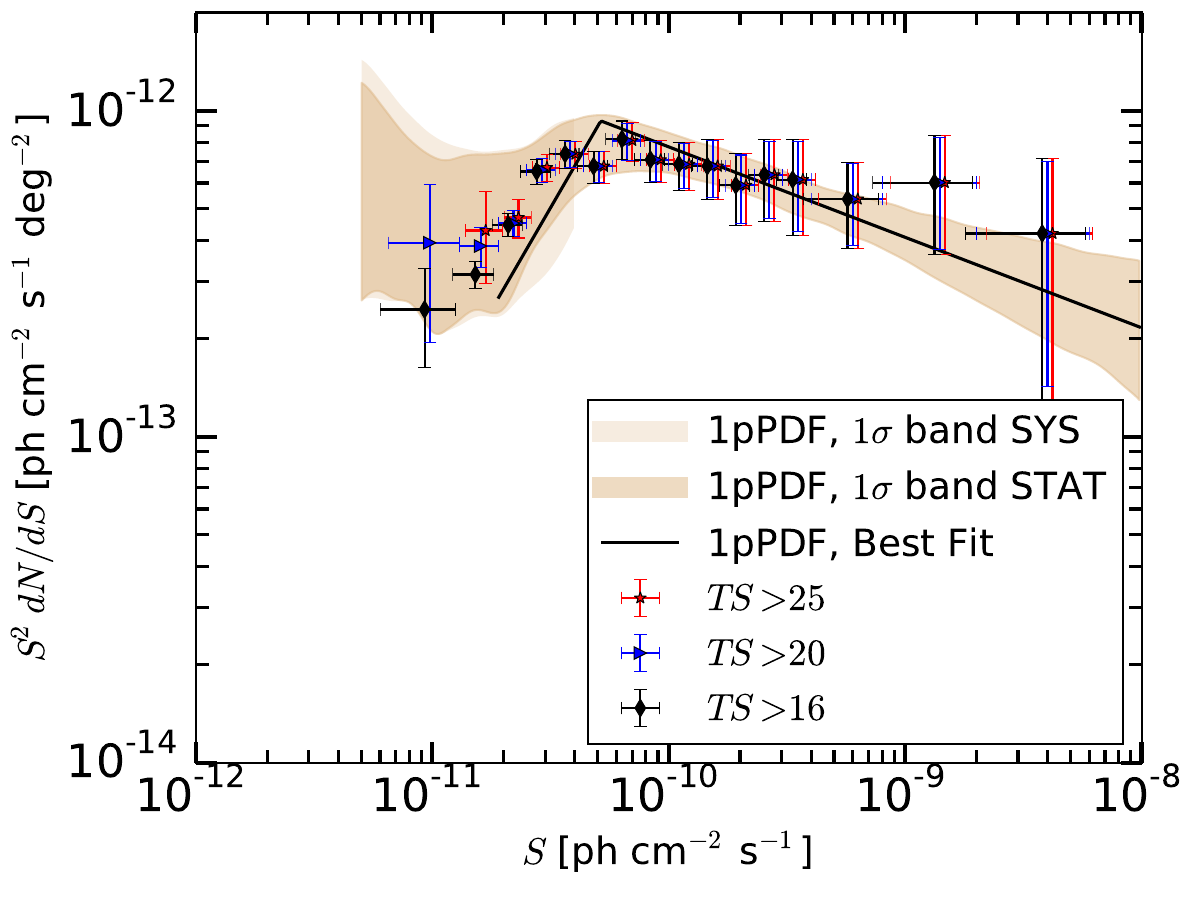}
\caption{Differential source-count distribution $dN/dS$ obtained with the 1pPDF method, using 7-year \textit{Fermi}-LAT data with {\tt UCV} event selection cuts. Same as the right panel of Figure~\ref{fig:dNdSfit}, but using pixel resolution order~8.}
\label{fig:1pPDF_hp8}
\end{centering}
\end{figure}

\end{appendix}

\bibliography{paper}

\begin{thebibliography}{34}
\expandafter\ifx\csname natexlab\endcsname\relax\def\natexlab#1{#1}\fi

\bibitem[{{Abdo} {et~al.}(2010){Abdo}, {Ackermann}, {Ajello}, {Antolini},
  {et~al.}}]{2010ApJ...720..435A}
{Abdo}, A.~A., {Ackermann}, M., {Ajello}, M., {Antolini}, E., {et~al.} 2010,
  \apj, 720, 435

\bibitem[{{Acero} {et~al.}(2015){Acero}, {Ackermann}, {Ajello},
  {et~al.}}]{2015ApJS..218...23A}
{Acero}, F., {Ackermann}, M., {Ajello}, M., {et~al.} 2015, \apjs, 218, 23

\bibitem[{Acero {et~al.}(2016)}]{Acero:2016qlg}
Acero, F. {et~al.} 2016, Astrophys. J. Suppl., 223, 26

\bibitem[{{Ackermann} {et~al.}(2017){Ackermann}, {Ajello}, {Albert},
  {et~al.}}]{FermiPass8GC}
{Ackermann}, M., {Ajello}, M., {Albert}, A., {et~al.} 2017

\bibitem[{{Ackermann} {et~al.}(2012){Ackermann}, {Ajello}, {Allafort},
  {et~al.}}]{2012ApJ...755..164A}
{Ackermann}, M., {Ajello}, M., {Allafort}, A., {et~al.} 2012, \apj, 755, 164

\bibitem[{Ackermann {et~al.}(2013)}]{Ackermann:2013fwa}
Ackermann, M. {et~al.} 2013, Astrophys. J. Suppl., 209, 34

\bibitem[{Ackermann {et~al.}(2014)}]{Fermi-LAT:2014sfa}
Ackermann, M. {et~al.} 2014, Astrophys. J., 793, 64

\bibitem[{Ackermann {et~al.}(2015)}]{Ackermann:2014usa}
Ackermann, M. {et~al.} 2015, Astrophys.J., 799, 86

\bibitem[{Ackermann {et~al.}(2016)}]{TheFermi-LAT:2015ykq}
Ackermann, M. {et~al.} 2016, Phys. Rev. Lett., 116, 151105

\bibitem[{Ajello {et~al.}(2015)Ajello, Gasparrini, S\'anchez-Conde, Zaharijas,
  Gustafsson, {et~al.}}]{Ajello:2015mfa}
Ajello, M., Gasparrini, D., S\'anchez-Conde, M., {et~al.} 2015, Astrophys.J.,
  800, L27

\bibitem[{Ajello {et~al.}(2012)Ajello, Shaw, Romani, Dermer, Costamante,
  {et~al.}}]{Ajello:2011zi}
Ajello, M., Shaw, M., Romani, R., {et~al.} 2012, \apj, 751, 108

\bibitem[{Ajello {et~al.}(2017{\natexlab{a}})}]{TheFermi-LAT:2017pvy}
Ajello, M. {et~al.} 2017{\natexlab{a}}, Astrophys. J. Suppl., 232, 18

\bibitem[{Ajello {et~al.}(2017{\natexlab{b}})}]{Fermi-LAT:2017yoi}
Ajello, M. {et~al.} 2017{\natexlab{b}}, Submitted to: Astrophys. J.

\bibitem[{Atwood {et~al.}(2013)Atwood, Baldini, Bregeon, Bruel, Chekhtman,
  {et~al.}}]{Atwood:2013dra}
Atwood, W., Baldini, L., Bregeon, J., {et~al.} 2013, Astrophys.J., 774, 76

\bibitem[{Bechtol {et~al.}(2017)Bechtol, Ahlers, Di~Mauro, Ajello, \&
  Vandenbroucke}]{Bechtol:2015uqb}
Bechtol, K., Ahlers, M., Di~Mauro, M., Ajello, M., \& Vandenbroucke, J. 2017,
  Astrophys. J., 836, 47

\bibitem[{{Casandjian} {et~al.}(2009){Casandjian}, {Grenier}, \& {for the Fermi
  Large Area Telescope Collaboration}}]{2009arXiv0912.3478C}
{Casandjian}, J.-M., {Grenier}, I., \& {for the Fermi Large Area Telescope
  Collaboration}. 2009, arXiv:0912.3478

\bibitem[{{Damiani} {et~al.}(1997){Damiani}, {Maggio}, {Micela}, \&
  {Sciortino}}]{1997ApJ...483..350D}
{Damiani}, F., {Maggio}, A., {Micela}, G., \& {Sciortino}, S. 1997, \apj, 483,
  350

\bibitem[{Di~Mauro {et~al.}(2014{\natexlab{a}})Di~Mauro, Calore, Donato,
  Ajello, \& Latronico}]{DiMauro:2013xta}
Di~Mauro, M., Calore, F., Donato, F., Ajello, M., \& Latronico, L.
  2014{\natexlab{a}}, \apj, 780, 161

\bibitem[{Di~Mauro {et~al.}(2014{\natexlab{b}})Di~Mauro, Cuoco, Donato, \&
  Siegal-Gaskins}]{DiMauro:2014wha}
Di~Mauro, M., Cuoco, A., Donato, F., \& Siegal-Gaskins, J.~M.
  2014{\natexlab{b}}, JCAP, 1411, 021

\bibitem[{Di~Mauro \& Donato(2015)}]{DiMauro:2015tfa}
Di~Mauro, M. \& Donato, F. 2015, Phys.Rev., D91, 123001

\bibitem[{Di~Mauro {et~al.}(2014{\natexlab{c}})Di~Mauro, Donato, Lamanna,
  Sanchez, \& Serpico}]{DiMauro:2013zfa}
Di~Mauro, M., Donato, F., Lamanna, G., Sanchez, D., \& Serpico, P.
  2014{\natexlab{c}}, Astrophys.J., 786, 129

\bibitem[{Eddington(1913)}]{eddington1913}
Eddington, A.~S. 1913, \mnras, 359, 73

\bibitem[{Fornasa \& S\'anchez-Conde(2015)}]{Fornasa:2015qua}
Fornasa, M. \& S\'anchez-Conde, M.~A. 2015, Phys. Rept., 598, 1

\bibitem[{Gavish \& Eichler(2016)}]{Gavish:2016tfl}
Gavish, E. \& Eichler, D. 2016, Astrophys. J., 822, 56

\bibitem[{{G{\'o}rski} {et~al.}(2005){G{\'o}rski}, {Hivon}, {Banday},
  {Wandelt}, {Hansen}, {Reinecke}, \& {Bartelmann}}]{2005ApJ...622..759G}
{G{\'o}rski}, K.~M., {Hivon}, E., {Banday}, A.~J., {et~al.} 2005, \apj, 622,
  759

\bibitem[{Lamastra {et~al.}(2017)Lamastra, Menci, Fiore, Antonelli,
  Colafrancesco, Guetta, \& Stamerra}]{Lamastra:2017iyo}
Lamastra, A., Menci, N., Fiore, F., {et~al.} 2017, Astron. Astrophys., 607, A18

\bibitem[{Lisanti {et~al.}(2016)Lisanti, Mishra-Sharma, Necib, \&
  Safdi}]{Lisanti:2016jub}
Lisanti, M., Mishra-Sharma, S., Necib, L., \& Safdi, B.~R. 2016, Astrophys. J.,
  832, 117

\bibitem[{Malyshev \& Hogg(2011)}]{Malyshev:2011zi}
Malyshev, D. \& Hogg, D.~W. 2011, Astrophys. J., 738, 181

\bibitem[{{Starck} \& {Pierre}(1998)}]{1998A&AS..128..397S}
{Starck}, J.-L. \& {Pierre}, M. 1998, \aaps, 128, 397

\bibitem[{{Su} {et~al.}(2010){Su}, {Slatyer}, \&
  {Finkbeiner}}]{2010ApJ...724.1044S}
{Su}, M., {Slatyer}, T.~R., \& {Finkbeiner}, D.~P. 2010, \apj, 724, 1044

\bibitem[{Wilks(1938)}]{wilks1938}
Wilks, S.~S. 1938, Ann. Math. Statist., 9, 60

\bibitem[{{Wood} {et~al.}(2017){Wood}, {Caputo}, {Charles}, {Di Mauro},
  {Magill}, \& {Jeremy Perkins for the Fermi-LAT
  Collaboration}}]{2017arXiv170709551W}
{Wood}, M., {Caputo}, R., {Charles}, E., {et~al.} 2017, ArXiv:1707.09551

\bibitem[{Zechlin {et~al.}(2016{\natexlab{a}})Zechlin, Cuoco, Donato, Fornengo,
  \& Regis}]{Zechlin:2016pme}
Zechlin, H.-S., Cuoco, A., Donato, F., Fornengo, N., \& Regis, M.
  2016{\natexlab{a}}, Astrophys. J., 826, L31

\bibitem[{Zechlin {et~al.}(2016{\natexlab{b}})Zechlin, Cuoco, Donato, Fornengo,
  \& Vittino}]{Zechlin:2015wdz}
Zechlin, H.-S., Cuoco, A., Donato, F., Fornengo, N., \& Vittino, A.
  2016{\natexlab{b}}, Astrophys. J. Suppl., 225, 18

\end{thebibliography}

\end{document}